\documentclass[12pt]{article}

\usepackage{amssymb}
\usepackage[dvips]{graphicx}
\makeatletter
 \@addtoreset{equation}{section}
 
\makeatother

\newif\ifContLineOne
\newif\ifContLineTwo
\newif\ifContLineThree

\def\conC#1{\vbox{\ialign{##\crcr
  \ifContLineThree\hrulefill\else\vphantom{\hrulefill}\fi\crcr
  \noalign{\kern3.2pt\nointerlineskip}
  \ifContLineTwo\hrulefill\else\vphantom{\hrulefill}\fi\crcr
  \noalign{\kern3.2pt\nointerlineskip}
  \ifContLineOne\hrulefill\else\vphantom{\hrulefill}\fi\crcr
  \noalign{\nointerlineskip}
  $\hfil\textstyle{\vbox to 14pt{}#1}\hfil$\crcr}}}

\def\DrawLeg#1#2{
  \kern-.2pt              % back up half width of leg
  \dimen2 =#1             % =height of whatever is underneath leg
  \advance\dimen2 by 2pt  % 2pt space below bottom of leg
  \dimen3 = 10.6pt        % base value of height of top of leg
  \dimen4 =3.6pt          % add this much time 1 2 or 3 to base value
  \advance\dimen3 by -\dimen2 
  \multiply\dimen4 by #2
  \advance\dimen3 by \dimen4
  \raise\dimen2 \hbox{\vrule height\dimen3 width .4pt} % draw it
  \kern-.2pt}             % and back up half width of line

\def\begC#1#2{\setbox0 =\hbox{$\textstyle{#2}$}
  \dimen0=.5\wd0 \dimen1=\ht0
  \conC{\hskip\dimen0}
  \count255=#1
  \ifnum\count255 =1 \ContLineOnetrue\else
  \ifnum\count255 =2 \ContLineTwotrue\else
  \ifnum\count255 =3 \ContLineThreetrue\fi\fi\fi
  \DrawLeg{\dimen1}{\count255}
  \conC{\hskip\dimen0}
  \kern-\dimen0\kern-\dimen0 \box0}

\def\endC#1#2{\setbox0 =\hbox{$\textstyle{#2}$}
  \dimen0=.5\wd0 \dimen1=\ht0
  \conC{\hskip\dimen0}
  \count255=#1
  \ifnum\count255 =1 \ContLineOnefalse\else
  \ifnum\count255 =2 \ContLineTwofalse\else
  \ifnum\count255 =3 \ContLineThreefalse\fi\fi\fi
  \DrawLeg{\dimen1}{\count255}
  \conC{\hskip\dimen0}
  \kern-\dimen0\kern-\dimen0 \box0}

\begin{document}

%%%%%%%%%%% Title Page %%%%%%%%%%%%%%%%%%%%

\begin{titlepage}

\renewcommand{\thefootnote}{\fnsymbol{footnote}}

\begin{flushright}
\begin{tabular}{l}
UTHEP-624\\
OIQP-11-03
\end{tabular}
\end{flushright}

\bigskip

\begin{center}
{\Large \bf 
Spacetime Fermions in Light-cone Gauge Superstring Field Theory
   and Dimensional Regularization
}
\end{center}

\bigskip

\begin{center}
%% AUTHORS
{\large Nobuyuki Ishibashi}${}^{a}$\footnote{e-mail:
        ishibash@het.ph.tsukuba.ac.jp}
and
{\large Koichi Murakami}${}^{b}$\footnote{e-mail:
        koichimurakami71@gmail.com}
\end{center}

\begin{center}
${}^{a}$\textit{Institute of Physics, University of Tsukuba,\\
Tsukuba, Ibaraki 305-8571, Japan}\\
\end{center}
\begin{center}
$^{b}$\textit{Okayama Institute for Quantum Physics,\\
      Kyoyama 1-9-1, Kita-ku, Okayama 700-0015, Japan} 
\end{center}

\bigskip

\bigskip

\bigskip

\begin{abstract}
We consider the dimensional regularization of the light-cone gauge 
type~II superstring field theories in the NSR formalism. 
In the previous work, we have calculated the tree-level amplitudes 
with external lines in the (NS,NS) sector using the regularization
and shown that the desired results are obtained without introducing
contact term interactions. 
In this work, we study the tree-level amplitudes with external lines 
in the Ramond sector. In order to deal with them,
we propose a worldsheet theory to be used instead of that for the
naive dimensional regularization. With the worldsheet theory, 
we regularize and define the tree-level amplitudes 
by analytic continuation.
We show that the results coincide with those of
the first quantized formulation. 
\end{abstract}

\setcounter{footnote}{0}
\renewcommand{\thefootnote}{\arabic{footnote}}

\end{titlepage}

%%%%%%%%%%%%%%%%%%%%%%%%%%%%%%%%%%%%%
\section{Introduction}

It is desirable to find a good way to regularize ultraviolet and infrared
divergences in string field theory. Although regularization did not
play essential roles in calculating scattering amplitudes, it will
be needed to deal with amplitudes which involve off-shell quantities
such as boundary states. In the superstring field theory, regularization
is needed to deal with the contact term 
problem~\cite{Greensite:1986gv,Greensite:1987hm,Greensite:1987sm,%%
Green:1987qu,Wendt:1987zh}.

In the previous works~\cite{Baba:2009kr,Baba:2009ns,Baba:2009fi,Baba:2009zm},
we have proposed to dimensionally regularize the light-cone gauge
string field 
theory~\cite{Mandelstam:1973jk,Kaku:1974zz,Kaku:1974xu,Cremmer:1974ej,%%
Mandelstam:1974hk,Sin:1988yf}
to deal with the divergences. We have shown that the light-cone gauge
string theory in noncritical dimensions corresponds to a conformal
gauge string theory in a Lorentz noninvariant background which preserves
the BRST symmetry on the worldsheet. This implies that the dimensional
regularization in the light-cone gauge string field theory preserves
the gauge symmetry of the string field. We have calculated the tree-level
amplitudes for the type~II superstring field theories in the NSR formalism 
using this regularization,
when the external lines are in the (NS,NS) sector. The results of
the first quantized formulation are reproduced without any need for
introducing the contact term interactions in the analytic continuation
$d\rightarrow10$ performed in the end of the calculation.

In this work, we would like to generalize our analysis to the amplitudes
with external lines in the Ramond sector. In Ref.~\cite{Ishibashi:2010nq},
we have discussed how we can deal with the Ramond sector for superstrings
in noncritical dimensions. It is possible to define BRST invariant
vertex operators for the Ramond sector and obtain the BRST invariant form
of the amplitudes using them. However, it has also been pointed out
that the naive dimensional regularization has a problem in dealing
with the type~II theories. 
Namely, it turns out that the regularized theory cannot
have spacetime fermions in the spectrum.

In this paper, we propose an alternative to the naive dimensional
regularization. We modify the worldsheet theory so that the theory
includes spacetime fermions and at the same time the divergences are
regularized. Using the worldsheet theory, we calculate the amplitudes
and analytically continue the results to $d=10$. We show that this
procedure yields tree-level amplitudes which coincide with those of
the first quantized formalism, without any need for adding the contact
term interactions.

This paper is organized as follows. 
In section~\ref{sec:Dimensional-regularization},
we present the worldsheet theory to be used for the regularization.
We show that it can be used to regularize the superstring theory with
spacetime fermions. 
In section~\ref{sec:Dimensionally-regularized-amplitudes},
we explain how to perform the analytic continuation of the amplitudes
to define them for $d=10$. We show that the results obtained coincide
with those of the first quantized formalism. 
Section~\ref{sec:Conclusions-and-discussions}
is devoted to conclusions and discussions. 
In particular, we discuss how we can apply our dimensional regularization 
scheme to deal with the divergences of the multi-loop amplitudes. 
In appendix~\ref{sec:SFTaction},
we present the action of the light-cone gauge superstring field theory
and the calculation of the amplitudes.
%In appendix~\ref{sec:example}, we  illustrate the manipulation
%in section~\ref{sec:Dimensionally-regularized-amplitudes}
%by using a four point tree-level amplitude as an example.

\section{Dimensional regularization\label{sec:Dimensional-regularization}}

In the previous works, we have considered dimensional regularization
of the light-cone gauge string field theory exactly as 
in the field theory for particles. 
It is possible to define the action in $d$ dimensional
spacetime if $d\geq2$ is an integer. We have calculated the amplitudes
perturbatively and analytically continued them as functions of $d$.
This procedure works for bosonic strings and for superstrings as long
as we deal with only fields in the (NS,NS) sector. However, as was
pointed out in Ref.~\cite{Ishibashi:2010nq}, such a naive procedure
does not work if one wants to incorporate spacetime fermions. Naive
dimensional continuation implies that the level-matching condition
for the (R,NS) sector becomes
\begin{equation}
\mathcal{N}+\frac{d-2}{16} 
=  \tilde{\mathcal{N}}\ ,
\label{levelmatching}
\end{equation}
 where $\mathcal{N}$ and $\tilde{\mathcal{N}}$ denote the left and
the right mode numbers of the light-cone gauge string state, and there
are no states satisfying it for general $d$. The same argument applies
to the (NS,R) sector. Therefore the dimensionally regularized theory
cannot have spacetime fermions in the spectrum and we cannot use it
to regularize the type~II superstring theories. It may be used to
deal with the type~$0$ theories.

In order to deal with spacetime fermions, we need to find an alternative
to the naive dimensional regularization. In the calculation presented
in Ref.~\cite{Baba:2009zm}, one can see that the contact
term divergences are regularized by the factor $e^{-\frac{d-2}{16}\Gamma}$
in the light-cone gauge amplitudes which comes from the conformal
anomaly.%
\footnote{The explicit form of $\Gamma$ is given in eq.(\ref{Gamma}).%
}  Therefore what matters in regularization is the Virasoro central
charge of the worldsheet theory. Instead of considering the theory
in $d$ dimensional spacetime, we can consider the light-cone gauge
worldsheet theory with the Virasoro central charge 
$\frac{3}{2}(d-2)$ to make the amplitudes finite. 
As will be discussed 
in section~\ref{sec:Conclusions-and-discussions},
in contrast to particle theory, the ultraviolet behavior of 
the multi-loop amplitudes in string theory is also affected 
by the central charge 
rather than the number of the momentum variables.

\subsection{Worldsheet theory}
For dimensional regularization, we need to make the Virasoro central
charge less than the critical one. This can be achieved by adding
a superconformal field theory with negative central charge to the
usual worldsheet theory of the transverse variables 
$X^{i},\psi^{i},\tilde{\psi}^{i}$
$(i=1,\ldots,8)$. 
Actually we have considered such a worldsheet theory in 
Ref.~\cite{Baba:2009kr}. 
As long as we are dealing with the (NS,NS) sector, 
the amplitudes essentially depends only on the value of 
the Virasoro central charge of the added superconformal field theory. 
However, the Ramond sector depends on the details of the theory. 
In the following,
we would like to show that it is possible to choose 
the superconformal field theory so that the difficulty mentioned above 
can be avoided and we can deal with the spacetime fermions.

The value of $\frac{d-2}{16}$ on the left hand side of 
eq.(\ref{levelmatching}) comes from 
the conformal dimension of the spin field, which depends on the 
number of the fermionic coordinates. 
%%The worldsheet theory for the naive dimensional regularization 
%%consists of free variables.
%%The Virasoro central charge and the number of the fermionic 
%%coordinates are all fixed by the spacetime dimensions. 
If we add a free superconformal field theory 
to the worldsheet theory for dimensional regularization, 
the Virasoro central charge and the number of the fermionic 
coordinates are both fixed by the spacetime dimensions.
By choosing an interacting superconformal field theory, 
one can change the Virasoro central charge and 
the dimension of the spin field independently. 
Then it would be possible to deal with the divergences 
and the level-matching condition simultaneously. 
An example of such a theory is the super WZW model. 
By changing the level, we can change the value of 
the Virasoro central charge 
keeping the dimension of the group manifold 
and therefore the dimension of the spin field. 
However, adding a super WZW model alone to the transverse variables 
is not enough, 
because the dimension of the spin field is still positive and 
the spectrum of the spacetime fermions becomes quite different from 
that for $d=10$. 
One can deal with this problem by adding variables corresponding to 
coordinates with opposite statistics, namely a ghost-like system. 
Such coordinates can be considered as those in the directions with 
``negative dimensions" and have the effect of reducing 
the dimension of the spin field.  
Therefore adding a superconformal field theory which consists of 
a super WZW model and a ghost-like system will do the job.  

In this paper, as an example of such a superconformal field theory,
we consider the one with the action
%As we explained above, as an example of
%superconformal field theories to be added 
%for dimensional regularization,
%we consider the one with the action
\begin{equation}
S=S_{G}+S_{\mathrm{gh}}\ ,
\end{equation}
 where
\begin{eqnarray}
S_{G} & = & 
  kS_{\mathrm{WZW}}\left[g\right]
  +\frac{1}{\pi}\int d^{2}z
      \left(\lambda^{a}\bar{\partial}\lambda^{a}
             +\tilde{\lambda}^{a}\partial\tilde{\lambda}^{a}
      \right)\ ,
\nonumber \\
S_{\mathrm{gh}} & = & 
  \frac{1}{\pi} \int d^{2}z
   \left(b^{A}\bar{\partial}c^{A}
         +\tilde{b}^{A}\partial\tilde{c}^{A}
         +\beta^{A}\bar{\partial}\gamma^{A}
         +\tilde{\beta}^{A}\partial\tilde{\gamma}^{A}
   \right)\ .
\label{eq:sWZW+sCC-}
\end{eqnarray}
 $S_{G}$ denotes the action of the super WZW model of level $k$
for the group $G=\left(SU(2)\right)^{2M}$,
where $M\geq0$ and $k>0$ are integers.
$S_{\mathrm{WZW}}\left[g\right]$ is the action 
for the $\left(SU(2)\right)^{2M}$ WZW model, 
$\lambda^{a}$ and $\tilde{\lambda}^{a}$ are the fermions
with weight $\left(\frac{1}{2},0\right),\left(0,\frac{1}{2}\right)$
respectively, and $a=1,\cdots,6M$ are the indices in the adjoint
representation of $\left(SU(2)\right)^{2M}$. The ghost-like variables
$b^{A},c^{A},$$\tilde{b}^{A},\tilde{c}^{A},$
$\beta^{A},\gamma^{A},$$\tilde{\beta}^{A},\tilde{\gamma}^{A}$
are with $A=1,\cdots,3M$. $b^{A},c^{A},\tilde{b}^{A},\tilde{c}^{A}$
are Grassmann odd variables with conformal weight $\left(1,0\right),$
$\left(0,0\right),$ $\left(0,1\right),$ $\left(0,0\right)$ respectively
and $\beta^{A},\gamma^{A},\tilde{\beta}^{A},\tilde{\gamma}^{A}$ are
Grassmann even and with conformal weight $\left(\frac{1}{2},0\right),$
$\left(\frac{1}{2},0\right),$ 
$\left(0,\frac{1}{2}\right),$ $\left(0,\frac{1}{2}\right)$
respectively. We take the boundary conditions of $\gamma^{A},\beta^{A}$
(resp.\ $\tilde{\gamma}^{A},\tilde{\beta}^{A}$) to be the same as
those of $\psi^{i},\lambda^{a}$ 
(resp.\ $\tilde{\psi}^{i},\tilde{\lambda}^{a}$).
Thus the theory is a superconformal field theory consisting of the
super WZW model and a ghost-like system. 

We take the worldsheet theory
to be that of the transverse coordinates $X^{i},\psi^{i},\tilde{\psi}^{i}$
combined with the CFT defined above. Therefore the total central charge
is
\begin{equation}
\begin{array}{ccccccccccc}
 &  &\mbox{\small $X^i,\psi^i$}  & & \mbox{\small $\mathrm{WZW}$}
  & & \mbox{\small $\lambda^{a}$} & & 
  \mbox{\small $b^{A},c^{A},\beta^{A},\gamma^{A}$} & &\\[0.75ex]
c & = & 12 &+& \displaystyle \frac{6Mk}{k+2} & +& 3M &+& (-9M)
  &=& \displaystyle 12-\frac{12M}{k+2}~,
\end{array}
\label{eq:c}
\end{equation}
 and we may regard the resultant worldsheet theory as the
 ``transverse part" of a theory in the effective spacetime dimension
\begin{equation}
d  \equiv  \frac{2}{3}c+2=10-\frac{8M}{k+2}\ .
\label{eq:d}
\end{equation}
 By varying $M$ with $k$ fixed, one can realize superconformal field
theories with largely negative central charge. For such theories, the
amplitudes are finite because of the factor $e^{-\frac{d-2}{16}\Gamma}$.
Using these theories, one can get an expression of the amplitudes
which can be analytically continued as a function of $M$, as will
be explained in section~\ref{sec:Dimensionally-regularized-amplitudes}.
The critical dimension is recovered in the limit $M\to0$.

\subsection{External lines}

We would like to write down the dimensionally regularized version
of the amplitudes using the modified worldsheet theory. In order to
do so, we consider the amplitudes with the external lines specified
in the following way. 
Suppose that for $d=10$ each external line corresponds to
a state $\left|\ \right\rangle _{X^{i},\psi^{i},\tilde{\psi}^{i}}$
in the Fock space of the transverse variables. We bosonize 
$\lambda^{a},\beta^{A},\gamma^{A}$
as 
\begin{eqnarray}
\frac{1}{\sqrt{2}}\left(\lambda^{2A-1}\pm i\lambda^{2A}\right) 
 & = & e^{\pm i\varphi^{A}}\ ,
\nonumber \\
\gamma^{A} 
 & = & \eta^{A}e^{\phi^{A}}\ ,
\nonumber \\
\beta^{A} & = & e^{-\phi^{A}}\partial\xi^{A}\ ,
\end{eqnarray}
 in the usual way so that we can express the vertex operators in the
Ramond sector. 
Here $\varphi^A,\phi^A$ are chiral bosons and 
$\eta^A,\xi^A$ are Grassmann odd variables with weight $(1,0),(0,0)$. 
As the regularized version for the amplitudes, we consider
those with the external lines 
\begin{eqnarray}
\left|\ \right\rangle _{X^{i},\psi^{i},\tilde{\psi}^{i}}
  \otimes\left|0\right\rangle _{L}
  \otimes\left|0\right\rangle _{R} 
  &  & \mathrm{(NS,NS)\, sector}\ ,
 \nonumber \\
\left|\ \right\rangle _{X^{i},\psi^{i},\tilde{\psi}^{i}}
  \otimes\left|\pm\right\rangle _{L}
  \otimes\left|0\right\rangle _{R}
  &  & \mathrm{(R,NS)\, sector}\ ,
 \nonumber \\
\left|\ \right\rangle _{X^{i},\psi^{i},\tilde{\psi}^{i}}
  \otimes\left|0\right\rangle _{L}
  \otimes\left|\pm\right\rangle _{R} 
  &  & \mathrm{(NS,R)\, sector}\ ,
 \nonumber \\
\left|\ \right\rangle _{X^{i},\psi^{i},\tilde{\psi}^{i}}
  \otimes\left|\pm\right\rangle _{L}
  \otimes\left|\pm\right\rangle _{R} 
  &  & \mathrm{(R,R)\, sector}\ ,
\label{eq:external}
\end{eqnarray}
 where $\left|0\right\rangle _{L,R}$ denote the left and right moving
$SL(2,\mathbb{C})$ invariant vacua of the superconformal field theory
defined above and $\left|\pm\right\rangle_L $ denote the states 
corresponding to the primary fields 
\begin{equation}
\prod_{A}e^{\pm\frac{1}{2}\phi^{A}}
\prod_{A}e^{\pm\frac{i}{2}\varphi^{A}}\ ,
\label{eq:externalprimary}
\end{equation}
 whose conformal weight is $0$. 
$\left|\pm\right\rangle_R $ are defined in the same way. 
Therefore the level-matching conditions
for the (R,NS) and (NS,R) sectors are the same as those for the transverse
part. Thus all the spacetime fermions in the critical theory have
their cousins in the regularized theory. 
Given any amplitudes for $d=10$, 
one can write down a regularized version by choosing the external lines 
from eq.(\ref{eq:external}) so that the amplitudes 
do not vanish identically. 
With this choice of
the external lines, we do not have any trouble in dealing with spacetime
fermions.

\section{Dimensionally regularized amplitudes
    \label{sec:Dimensionally-regularized-amplitudes}}

Let us consider the tree-level amplitudes 
of the light-cone gauge string field theory 
corresponding to the worldsheet theory defined in the previous section. 
We obtain them as functions of $M$, with $M$ non-negative integer. 
For $M$ big enough, the divergences are regularized 
and the amplitudes are well-defined.
We would like to define the amplitudes as analytic functions of $M$
and obtain the amplitudes for $d=10$ by taking the limit $M\to0$.
Since the amplitudes are given only for non-negative integer $M$,
we should specify the way to preform 
the analytic continuation.\footnote{ 
This problem was not discussed in the previous 
works~\cite{Baba:2009kr,Baba:2009zm}, 
in which we considered amplitudes with (NS,NS) external lines. 
If all the external lines are in the (NS,NS) sector, 
it is possible to actually realize the amplitudes 
for noninteger $d$ by choosing the worldsheet theory with 
the Virasoro central charge $c=\frac{3}{2}(d-2)$.  
Then the analytic continuation is obvious.}

\subsection{Light-cone gauge amplitudes}

The string field theory action and the calculation of the amplitudes
are explained in appendix~\ref{sec:SFTaction}. The regularized tree-level
amplitudes for $N$ strings can be expressed as an integral over the
moduli space of the string diagram: 
\begin{equation}
\mathcal{A}_{N}
  = \left(4ig\right)^{N-2}
    \int \left( \prod_{\mathcal{I}=1}^{N-3}
                 \frac{d^{2}\mathcal{T}_{\mathcal{I}}}{4\pi}
         \right)
    F_{N}\left(\mathcal{T}_{\mathcal{I}},
               \bar{\mathcal{T}}_{\mathcal{I}}\right),
\label{eq:AN-1}
\end{equation}
where  
$\mathcal{T}_{\mathcal{I}},\bar{\mathcal{T}}_{\mathcal{I}}$
$(\mathcal{I}=1,\ldots,N-3)$
are the moduli parameters, which correspond to the Schwinger parameters
for the propagators in each channel. 
The integrand 
$F_{N}\left(\mathcal{T}_{\mathcal{I}},
            \bar{\mathcal{T}}_{\mathcal{I}}\right)$
is described by using the correlation function 
$\left\langle \cdots\right\rangle _{\mathrm{LC}}$
for the light-cone gauge worldsheet theory given in the last section
as 
\begin{eqnarray}
F_{N}\left(\mathcal{T}_{\mathcal{I}},
           \bar{\mathcal{T}}_{\mathcal{I}}\right)
 & = & \left(2\pi\right)^{2}
     \delta\left(\sum_{r=1}^{N}p_{r}^{+}\right)
     \delta\left(\sum_{r=1}^{N}p_{r}^{-}\right)
     \mathrm{sgn}\left(\prod_{r=1}^{N}\alpha_{r}\right)
     e^{-\frac{d-2}{16}\Gamma}
     f\left(\alpha_{r};Z_{r}\right)
\nonumber \\
&  & \ \times
  \left\langle 
      \prod_{A}\left(c^{A}\left(z_{0}\right)
              \tilde{c}^{A}\left(\bar{z}_{0}\right)\right)
      \prod_{I=1}^{N-2}
           \left|\left(\partial^{2}\rho\right)^{-\frac{3}{4}}
                 T_{F}^{\mathrm{LC}}\left(z_{I}\right)\right|^{2}
      \prod_{r=1}^{N}V_{r}^{\mathrm{LC}}
  \right\rangle _{\mathrm{LC}},
\label{eq:LCFN}
\end{eqnarray}
where  the function $f$ is given in eq.(\ref{eq:f}),  
$T_{F}^{\mathrm{LC}}$ denotes the supercurrent
of the worldsheet theory,
and $V_{r}^{\mathrm{LC}}$ are the vertex operators corresponding
to the states in eq.(\ref{eq:external}).
$V_{r}^{\mathrm{LC}}$ can be obtained
by combining the left and right contributions which take the form 
given in Ref.~\cite{Ishibashi:2010nq} but this time
they involve the operators in eq.(\ref{eq:externalprimary}) from 
the superconformal field theory defined in 
section~\ref{sec:Dimensional-regularization}. 
% The correlation function in eq.(\ref{eq:LCFN})
%can be expressed as that on the complex $z$-plane, via the Mandelstam
%mapping 
$\rho (z)$ is the Mandelstam mapping defined as
\begin{equation}
\rho(z)=\sum_{r=1}^{N}\alpha_{r}\ln(z-Z_{r})~,
\label{eq:Mandelstam}
\end{equation}
which maps the complex $z$-plane to the $N$-string
light-cone diagram parametrized by the complex coordinate $\rho$
as usual. $z_{I}$ ($I=1,\ldots,N-2$) are the interaction points
determined by $\partial\rho(z_{I})=0$. $\Gamma$ is defined by 
\begin{equation}
e^{-\Gamma}
  =\left| \sum_{r=1}^{N}\alpha_{r}Z_{r} \right|^{4}
   \prod_{r=1}^{N} \left(|\alpha_{r}|^{-2}
                    e^{-2\mathop{\mathrm{Re}}\bar{N}_{00}^{rr}}
                   \right)
   \prod_{I=1}^{N-2}\left|\partial^{2}\rho(z_{I})\right|^{-1}~,
\label{Gamma}
\end{equation}
 where $\bar{N}_{00}^{rr}$ is a Neumann coefficient given as 
\begin{equation}
\bar{N}_{00}^{rr}
  =\frac{\tau_{0}^{(r)}+i\beta_{r}}{\alpha_{r}}
    -\sum_{s\neq r}\frac{\alpha_{s}}{\alpha_{r}}\ln(Z_{r}-Z_{s})~,
\qquad \tau_{0}^{(r)}+i\beta_{r} \equiv\rho(z_{I^{(r)}})~.
\end{equation}
 Here $z_{I^{(r)}}$ denotes the interaction point where the $r$-th
string interacts. The amplitudes (\ref{eq:AN-1}) are defined for
non-negative integer $M$.

Let us see that by taking $M$ large enough, 
one can regularize the divergences
in the amplitudes
caused by the colliding supercurrents $T^{\mathrm{LC}}_{F} (z_{I})$
inserted at the interaction points,
$T^{\mathrm{LC}}_{F} (z_{I}) T^{\mathrm{LC}}_{F} (z_{J})
   \sim (z_{I}-z_{J})^{-3}$.
Because of the identity
\begin{equation}
\partial^{2} \rho (z_{I})
  =\left( \sum_{s=1}^{N} \alpha_{s} Z_{s} \right)
    \frac{\prod_{J \neq I} (z_{I}-z_{J})}
         {\prod_{r=1}^{N} (z_{I} - Z_{r})}~,
\label{eq:Gamma-zIzJ}
\end{equation}
one can find that $e^{-\frac{d-2}{16}\Gamma}$
obtained from eq.(\ref{Gamma}) behaves as
$e^{-\frac{d-2}{16}\Gamma}
 \sim \left| z_{I} - z_{J} \right|^{-\frac{d-2}{8}}$
for $z_{I} \sim z_{J}$.
It follows that 
the most divergent term in the integrand $F_{N}$
in eq.(\ref{eq:LCFN}) behaves as
\begin{equation}
e^{-\frac{d-2}{16}\Gamma} 
  \prod_{I} \left| (\partial^{2}\rho)^{-\frac{3}{4}} 
                   T_{F}^{\mathrm{LC}}(z_{I}) 
            \right|^{2}
\sim \left|z_{I} -z_{J} \right|^{-9-\frac{d-2}{8}}
\end{equation}
for $z_{I} \sim z_{J}$.
Therefore if we take $d<-70$, the integrand on the right hand side of eq.(\ref{eq:AN-1}) 
does not diverge but vanishes for $z_I=z_J$.
Due to eq.(\ref{eq:d}), this implies that if $M > 10(k+2)$, 
the divergences are indeed regularized.

Therefore we can define $\mathcal{A}_{N}$ for $M$ large enough and 
may analytically continue it to $M=0$.  In order to do so, 
we should specify the way to define $\mathcal{A}_{N}$ as an analytic
function of $M$. It is easy to see that the $M$ dependence of $F_{N}$
arises in the following way: 
\begin{itemize}
\item Through the total central charge $c$ in eq.(\ref{eq:c}) 
 which appears in various places. 
 One is in the anomaly contribution to the amplitudes
\begin{equation}
e^{-\frac{d-2}{16}\Gamma} 
 =  e^{-\frac{1}{2}\Gamma+M\frac{\Gamma}{2\left(k+2\right)}}\ .
\end{equation}
 The on-shell condition also depends on $c$. 
 One can see that each
 vertex operator $V_{r}^{\mathrm{LC}}$ involves a factor
\begin{equation}
\exp \left(-\frac{M}{2\left(k+2\right)}
           \frac{\tau_{0}^{\left(r\right)}}{p_{r}^{+}}\right)\ ,
\end{equation}
 which comes from the on-shell condition. 
\item Since the superconformal field theory defined in 
 section~\ref{sec:Dimensional-regularization}
 consists of $M$ copies of $SU(2)^{2}$ super WZW model and ghost-like 
 system, the correlation functions of the theory depend on $M$. With
 the external lines given in eq.(\ref{eq:external}), the correlation
 functions on the complex plane depend on $M$ through the combinatorial
 factors, which can be expressed as a polynomial of $M$. 
\end{itemize}
Altogether we can see that $F_{N}$ is given in the form 
\begin{equation}
F_{N}=P_N\left(M\right)e^{MQ_N}\ ,
\label{eq:FN-1}
\end{equation}
 where 
\begin{equation}
Q_N \equiv 
  \frac{1}{2\left(k+2\right)}
  \left(\Gamma
        -\sum_{r=1}^N \frac{\tau_{0}^{\left(r\right)}}
                           {p_{r}^{+}}
  \right)\ ,
\label{eq:def-QN}
\end{equation}
 and $P_N\left(M\right)$ is a polynomial of $M$. Thus we define $F_{N}$
for noninteger $M$ by analytically continuing the right hand side
of eq.(\ref{eq:FN-1}) and obtain $\mathcal{A}_{N}$ as an analytic
function of $M$.

\subsection{BRST invariant expression of the amplitudes 
  \label{sucsec:BRST-invariant-expression-of-the-amplitudes}}

It is obvious that in the limit $M\to0$, $\mathcal{A}_{N}$ 
in eq.(\ref{eq:AN-1})
coincides with the expression of the amplitudes for $d=10$. We define
the amplitudes $\mathcal{A}_{N}$ when $M$ is large enough, analytically
continue them and in the end of the calculation take the limit $M\to0$.
We would like to prove that the results are finite and coincide with
those of the first quantized formalism. In order to do so, we first
rewrite the light-cone gauge amplitude (\ref{eq:AN-1}) into the conformal
gauge one. Using the BRST invariance of the conformal gauge expression,
we can prove the above mentioned facts. Since the arguments are rather
lengthy, we first illustrate the proof for a simple case and treat
the general case later.

\subsubsection*{Four point amplitudes %Example of the manipulation in 
% subsection~\ref{sucsec:BRST-invariant-expression-of-the-amplitudes}
\label{sec:example}}

\begin{figure}[h]
\begin{centering}
\includegraphics[scale=0.7]{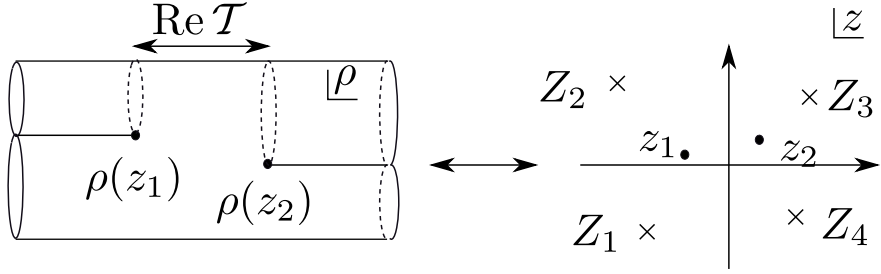} 
\par\end{centering}

\caption{Four point tree-level amplitudes\label{fig:Four-point-tree}}
\end{figure}

Let us consider the four point tree-level amplitudes with two external
lines in the (R,NS) sector and two in the (NS,NS) sector. The regularized
amplitude is given as 
\begin{equation}
\mathcal{A}_{4}
 = \left(4ig\right)^{2}
   \int\frac{d^{2}\mathcal{T}}{4\pi}
      F_{4}\left(\mathcal{T},\bar{\mathcal{T}}\right)~,
\end{equation}
 where the integrand $F_{4}\left(\mathcal{T},\bar{\mathcal{T}}\right)$
is described by using the correlation function as 
\begin{eqnarray}
F_{4}\left(\mathcal{T},\bar{\mathcal{T}}\right) 
 & = & \left(2\pi\right)^{2}
       \delta\left(\sum_{r=1}^{4}p_{r}^{+}\right)
       \delta\left(\sum_{r=1}^{4}p_{r}^{-}\right)
       \mathrm{sgn}\left(\prod_{r=1}^{4}\alpha_{r}\right)
       e^{-\frac{d-2}{16}\Gamma}
       f\left(\alpha_{r};Z_{r}\right)
\nonumber \\
 &  & \ \times
    \left\langle 
       \prod_{A}\left(c^{A}\left(z_{0}\right)
                      \tilde{c}^{A}\left(\bar{z}_{0}\right)\right)
       \prod_{I=1}^{2} \left|
           \left(\partial^{2}\rho\right)^{-\frac{3}{4}}
           T_{F}^{\mathrm{LC}}\left(z_{I}\right)
                      \right|^{2}
       \prod_{r=1}^{4} V_{r}^{\mathrm{LC}}
    \right\rangle _{\mathrm{LC}}.
\label{eq:F4LC}
\end{eqnarray}
 Here we have used the Mandelstam mapping $\rho\left(z\right)$ depicted
in Fig.~\ref{fig:Four-point-tree} to express the correlation function.
The complex moduli parameter $\mathcal{T}$ is given as 
\begin{equation}
\mathcal{T}=\rho\left(z_{2}\right)-\rho\left(z_{1}\right)\,.
\end{equation}
Let us consider the case where 
$V_{1}^{\mathrm{LC}}, V_{2}^{\mathrm{LC}}$
are in the (R,NS) sector. 

\subsubsection*{Conformal gauge expression}

We would like to express the correlation function in 
eq.(\ref{eq:F4LC}) using the conformal gauge variables. 
In the conformal gauge, the worldsheet
theory consists of the light-cone gauge one given in 
section~\ref{sec:Dimensional-regularization}
combined with the $X^{\pm}$ CFT~\cite{Baba:2009fi} and 
the super-reparametrization ghosts. 
With all the field content, one can define the nilpotent BRST
charge $Q_{\mathrm{B}}$. In order to rewrite eq.(\ref{eq:F4LC}),
we need the following identity: 
\begin{eqnarray}
\lefteqn{
  (2\pi)^{2}\delta\left(\sum_{r=1}^{4}p_{r}^{+}\right)
   \delta\left(\sum_{r=1}^{4}p_{r}^{-}\right)
   e^{-\frac{d-2}{16}\Gamma}
   \prod_{I=1}^{2} \left|\partial^{2}\rho\left(z_{I}\right)
                     \right|^{-\frac{3}{2}}
   f\left(\alpha_{r};Z_{r}\right)
   \prod_{r=1}^{4}V_{r}^{\mathrm{LC}}
}\nonumber \\
 &\sim& 
  \left\langle 
    \left|\partial\rho c(\infty)\right|^{2}
    \prod_{I=1}^{2}\left|\oint_{z_{I}}\frac{dz}{2\pi i}
                         \frac{b}{\partial\rho}(z)e^{\phi}(z_{I})
                   \right|^{2}
    \prod_{r=1}^{4}\mathcal{S}_{r}^{-1}
\right.
\nonumber \\
 &  & \hphantom{\langle\quad}
 \times\left.
   \vphantom{\prod_{I=1}^{2}}
   V_{1}^{\left(-\frac{3}{2},-1\right)}(Z_{1},\bar{Z}_{1})
   V_{2}^{\left(-\frac{1}{2},-1\right)}(Z_{2},\bar{Z}_{2})
   V_{3}^{\left(-1,-1\right)}(Z_{3},\bar{Z}_{3})
   V_{4}^{\left(-1,-1\right)}(Z_{4},\bar{Z}_{4})
\right\rangle _{X^{\pm},\mathrm{ghosts}}
\nonumber \\
 &  & \ \ \times \alpha_{1}^{\frac{1}{2}}\alpha_{2}^{\frac{1}{2}}~,
\label{eq:LCconformal}
\end{eqnarray}
where the $\sim$ means that the left
and right hand sides coincide up to an overall numerical constant.
Since such overall constants are irrelevant to show the BRST invariance,
we will henceforth ignore them.
 On the right hand side, the expectation value is taken with respect
to the $X^{\pm}$~CFT and the super-reparametrization ghosts. 
$V_{r}^{\left(p_{L,r},p_{R,r}\right)}$
denotes the vertex operator in the picture 
$\left(p_{L,r},p_{R,r}\right)$,
which can be obtained by combining the left and right ones defined
in Ref.~\cite{Ishibashi:2010nq}. 
$\mathcal{S}_{r}^{-1}$ is given in terms of
superfields~\cite{Ishibashi:2010nq} as
\begin{eqnarray}
\mathcal{S}_{r}^{-1}
 \equiv \oint_{z_{I^{(r)}}} \frac{d\mathbf{z}}{2\pi i}
        D\Phi (\mathbf{z})
        \oint_{\bar{z}_{I^{(r)}}} \frac{d\bar{\mathbf{z}}}{2\pi i}
        \bar{D} \Phi (\bar{\mathbf{z}})
        \, e^{\frac{d-10}{16} \frac{i}{p^{+}_{r}} \mathcal{X}^{+}}
          (\mathbf{z},\bar{\mathbf{z}})~,
\end{eqnarray}
where
\begin{eqnarray}
&&
\Phi  \equiv  
   \ln \left( -4\left(D\Theta^{+}\right)^{2}
               \left(\bar{D}\tilde{\Theta}^{+}\right)^{2}\right)\,,
\qquad
\Theta^{+}\left(\mathbf{z}\right) 
  \equiv  \frac{D\mathcal{X}^{+}}
                 {\left(\partial\mathcal{X}^{+}
                  \right)^{\frac{1}{2}}}
           \left(\mathbf{z}\right)\,,
\nonumber \\
&&
\mathcal{X}^{+}\left(\mathbf{z},\bar{\mathbf{z}}\right) 
  = X^{+} \left(z,\bar{z}\right)
        +i\theta\psi^{+} \left(z\right)
        +i\bar{\theta}\tilde{\psi}^{+} \left(\bar{z}\right)
        +i\theta\bar{\theta}F^{+} \left(z,\bar{z}\right)\,.
\end{eqnarray}
The identity~(\ref{eq:LCconformal}) was proved 
in Ref.~\cite{Ishibashi:2010nq} when all
the external lines are spacetime bosons. Actually it can be proved
in almost the same way even if spacetime fermions are involved. 
The only subtlety is that with spacetime fermions 
the superghost correlation function becomes left-right asymmetric 
and in the course of the calculation
we encounter phase factors which are constant  
but depend on how we take the cuts to
define the correlation function. We can proceed as in 
Ref.~\cite{Ishibashi:2010nq}
but obtain eq.(\ref{eq:LCconformal}) only up to such a phase factor.
However, in the final form~(\ref{eq:LCconformal}) we can see that
the phases add up to become just a numerical constant, because neither
the left nor the right hand side has cuts as a function of 
$Z_{r},\bar{Z}_{r},z_{I},\bar{z}_{I}$.
Thus eq.(\ref{eq:LCconformal}) holds also in our case.

Substituting eq.(\ref{eq:LCconformal}) into eq.(\ref{eq:F4LC}) we
obtain the conformal gauge expression of the amplitude: 
\begin{eqnarray}
 &  & \mathcal{A}_{4}
   \sim \int d^{2}\mathcal{T}
    \left\langle \left|\partial\rho c\left(\infty\right)\right|^{2}
        \prod_{A} \left(c^{A}\left(z_{0}\right)
                        \tilde{c}^{A}\left(\bar{z}_{0}\right)\right)
    \right.\nonumber \\
 &  & \hphantom{\mathcal{A}_{4}\sim\int d^{2}\mathcal{T}\langle}
  \left.\ \times
     \prod_{I=1}^{2} \left|\oint_{z_{I}}\frac{dz}{2\pi i}
                           \frac{b}{\partial\rho}\left(z\right)
                           e^{\phi}T_{F}^{\mathrm{LC}}
                                    \left(z_{I}\right)
                     \right|^{2}
     \prod_{r=1}^{4} \left[\mathcal{S}_{r}^{-1}
                           V_{r}^{\left(p_{L,r},p_{R,r}\right)}\right]
  \right\rangle 
  \alpha_{1}^{\frac{1}{2}}\alpha_{2}^{\frac{1}{2}}\,.~~~~~
\label{eq:confTF}
\end{eqnarray}
 Here $\left\langle \cdots\right\rangle $ denotes the correlation
function for the conformal gauge worldsheet theory. 
In eq.(\ref{eq:confTF}),
we can replace $e^{\phi}T_{F}^{\mathrm{LC}}$ 
by the picture changing operator 
\begin{equation}
X\left(z\right)\equiv\left\{ Q_{\mathrm{B}},\xi\left(z\right)\right\} \,,
\end{equation}
as was proved in Ref.~\cite{Ishibashi:2010nq}. 
The factor $\alpha_{1}^{\frac{1}{2}}\alpha_{2}^{\frac{1}{2}}$
is attributed to the normalization of the wave function for the
fermionic fields and will be ignored in the following. 
Thus we eventually obtain 
\begin{eqnarray}
& & \mathcal{A}_{4}
  \sim \int d^{2}\mathcal{T}
     \left\langle 
       \left|\partial\rho c\left(\infty\right)\right|^{2}
       \prod_{A}\left(c^{A}\left(z_{0}\right)
                      \tilde{c}^{A}\left(\bar{z}_{0}\right)
                \right)
     \right.
\nonumber \\
 &  & \hphantom{\mathcal{A}_{4}\sim\int d^{2}\mathcal{T}\langle}
 \left.\ \times\prod_{I=1}^{2}
               \left|\oint_{z_{I}}\frac{dz}{2\pi i}
                \frac{b}{\partial\rho}\left(z\right)
                X\left(z_{I}\right)\right|^{2}
           \prod_{r=1}^{4}  
                \left[\mathcal{S}_{r}^{-1}
                      V_{r}^{\left(p_{L,r},p_{R,r}\right)}
                \right]
 \right\rangle .
\label{eq:A4}
\end{eqnarray}

\begin{figure}
\begin{centering}
\includegraphics[scale=0.7]{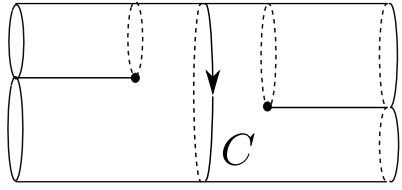} 
\par\end{centering}

\caption{Contour $C$\label{fig:Contour}}
\end{figure}

One can show that the expression on the right hand side of 
eq.(\ref{eq:A4}) is BRST invariant, 
if $M$ is large enough. Deforming the contour of 
  $\oint_{z_{I}}\frac{dz}{2\pi i}
       \frac{b}{\partial\rho}\left(z\right)$,
one easily gets the expression 
\begin{equation}
\mathcal{A}_{4}
 \sim  \int d^{2}\mathcal{T}
 \left\langle 
      \prod_{A}\left(c^{A}\left(z_{0}\right)
                     \tilde{c}^{A}\left(\bar{z}_{0}\right) \right)
     \left|\oint_{C}\frac{dz}{2\pi i}
                    \frac{b}{\partial\rho}\left(z\right)
     \right|^{2}
     \prod_{I=1}^{2} \left|X\left(z_{I}\right)\right|^{2}
     \prod_{r=1}^{4} \left[\mathcal{S}_{r}^{-1}
                            V_{r}^{\left(p_{L,r},p_{R,r}\right)}
                      \right]
 \right\rangle ,
\label{eq:A41}
\end{equation}
where $C$ is the contour depicted in Fig.~\ref{fig:Contour}. By
using the ``small Hilbert space" mentioned at the
end of appendix~\ref{sec:SFTaction}, 
the integrand on 
the right hand side of 
eq.(\ref{eq:A41}) is written as 
\begin{equation}
\left\langle 
   \left| \oint_{C}\frac{dz}{2\pi i}
                   \frac{b}{\partial\rho}\left(z\right)
   \right|^{2}
   \prod_{I=1}^{2} \left| X\left(z_{I}\right) \right|^{2}
   \prod_{r=1}^{4} \left[\mathcal{S}_{r}^{-1}
                         V_{r}^{\left(p_{L,r},p_{R,r}\right)}
                   \right]
\right\rangle _{\mathrm{small~Hilbert~space}}\,.
\end{equation}
%Since the operators we consider are such operators, 
%$\prod_{A}\left(c^{A}\left(z_{0}\right)
%\tilde{c}^{A}\left(\bar{z}_{0}\right)\right)$
In this form, it is easy to show the BRST invariance.%
\footnote{%%
In the large Hilbert space expression, the BRST variation of 
$\prod_{A}\left(c^{A}\left(z_{0}\right)
                \tilde{c}^{A}\left(\bar{z}_{0}\right)\right)$
does not vanish. However, the correlation functions including it vanish
if the other insertions are in the ``small Hilbert space". %
} We note that 
\begin{equation}
\left\{ Q_{\mathrm{B}}\,,
        \,\oint_{C}\frac{dz}{2\pi i}\frac{b}{\partial\rho}(z)
\right\} 
=\oint_{C}\frac{dz}{2\pi i}\frac{T}{\partial\rho}\left(z\right)\,.
\end{equation}
 The right hand side coincides with $L_{0}$ on the light-cone diagram
and therefore 
\begin{equation}
\left\{ Q_{\mathrm{B}}\,,
        \,\oint_{C}\frac{dz}{2\pi i}\frac{b}{\partial\rho}(z)
\right\} 
\sim -\partial_{\mathcal{T}}\,.
\label{eq:partialT}
\end{equation}
 If $M$ is large enough, we can ignore the surface terms and 
the total derivative terms vanish\footnote{
For $z_I\to z_J$, 
the factor $e^{-\frac{d-10}{16}\Gamma}$ which comes from 
the $X^\pm$ CFT dominates the divergent powers of $|z_I-z_J|$, 
if we take $-d$ to be large enough. 
Other surface terms are dealt with by analytically continuing 
the external momenta, as is usually done. 
}.
Since
the other insertions are BRST invariant, the right
hand side of eq.(\ref{eq:A4}) is BRST invariant with large enough
$M$.

The BRST invariant expression (\ref{eq:A4}) can also be defined for
noninteger $M$. The integrand on the right hand side of eq.(\ref{eq:A41})
coincides with $F_{4}$ in eq.(\ref{eq:F4LC}) and can be written
as eq.(\ref{eq:FN-1}). Therefore we analytically continue it
in terms of $M$ as in the previous subsection. We would like to show
that the limit $M\to0$ of the analytically continued amplitudes are
finite and coincide with those in the first quantized formalism. 
To do so, we first consider eq.(\ref{eq:A41}) for non-negative integer $M$ 
and move the picture changing operators 
$X(z_{I}),\tilde{X}(\bar{z}_{I})$
to the positions of external lines as was done in 
Ref.~\cite{Friedan:1985ge}.
Using eq.(\ref{eq:partialT}), we obtain%
\footnote{ Of course, such a manipulation is not valid if the external lines
are off-shell. It is possible to define dimensionally regularized
off-shell amplitudes using eq.(\ref{eq:A4}), but they cannot be written
in the form~(\ref{eq:A4finite}). %
} 
\begin{eqnarray}
 &  & 
\mathcal{A}_{4}
  \sim \int d^{2}\mathcal{T}
    \left[ 
      \left\langle 
           \prod_{A}\left(c^{A}\left(z_{0}\right)
                          \tilde{c}^{A}\left(\bar{z}_{0}\right)
                    \right)
          \left|\oint_{C}\frac{dz}{2\pi i}
                  \frac{b}{\partial\rho}\left(z\right)\right|^{2}
          \prod_{r=1}^{4}
            \left[\mathcal{S}_{r}^{-1}
                  V_{r}^{\left(p_{L,r}^{\prime},p_{R,r}^{\prime}\right)}
            \right]
      \right\rangle 
   \right.\nonumber \\
 &  & 
    \hphantom{\mathcal{A}_{4}\sim\int d^{2}\mathcal{T}\quad}
   {}+\partial_{\mathcal{T}}
   \left\langle
       \prod_{A}\left(c^{A}\left(z_{0}\right)
                      \tilde{c}^{A}\left(\bar{z}_{0}\right)
                \right)
     \oint_{C}\frac{d\bar{z}}{2\pi i}
              \frac{\tilde{b}}{\bar{\partial}\rho}(\bar{z})
     \tilde{X}(\bar{z}_{1})
     \left|X(z_{2})\right|^{2}
   \right.
\nonumber \\
&& \left. 
    \hphantom{\mathcal{A}_{4}\sim\int d^{2}\mathcal{T}
              + \partial_{\mathcal{T}} \langle \qquad}
\times
\prod_{r=1}^{4}\left[\mathcal{S}_{r}^{-1}
                     V_{r}^{\left(p_{L,r},p_{R,r}\right)}
               \right]
\right\rangle \nonumber \\
 &  & \hphantom{\mathcal{A}_{4}\sim\int d^{2}\mathcal{T}\quad}
{}+\left.\cdots
    \vphantom{\left\langle \left|\oint_{C}
              \frac{dz}{2\pi i}\frac{b}{\partial\rho}\left(z\right)
              \right|^{2}\right\rangle }
 \right]\,.
\label{eq:A42}
\end{eqnarray}
As a result of this manipulation,
the pictures of the vertex operators applied by picture changing operators
are shifted to be $\left(p_{L,r}^{\prime},p_{R,r}^{\prime}\right)$,
up to total derivative terms. The ellipses on the right hand side
denote the other total derivative terms which appear in the course
of the deformation of the contours of the BRST currents. For $M$ large
enough, the total derivative terms do not contribute to the integral
and the right hand side of eq.(\ref{eq:A41}) is equal to 
\begin{equation}
\int d^{2}\mathcal{T}
  \left\langle \prod_{A}
         \left(c^{A}\left(z_{0}\right)
               \tilde{c}^{A}\left(\bar{z}_{0}\right)
         \right)
         \left|\oint_{C}\frac{dz}{2\pi i}
               \frac{b}{\partial\rho}\left(z\right)
         \right|^{2}
       \prod_{r=1}^{4}
            \left[\mathcal{S}_{r}^{-1}
                  V_{r}^{\left(p_{L,r}^{\prime},p_{R,r}^{\prime}
                         \right)}
            \right]
\right\rangle \ .
\label{eq:A4finite}
\end{equation}
 In this form, the amplitudes are finite and coincide with those of
the first quantized formalism when $M=0$. The problem is whether
we can show eq.(\ref{eq:A42}) and express $\mathcal{A}_{4}$ 
as eq.(\ref{eq:A4finite})
even for noninteger $M$.

\subsubsection*{Amplitudes for noninteger $M$}

In order to define eqs.(\ref{eq:A42}) and (\ref{eq:A4finite}) 
for noninteger $M$, 
we would like to show that all the correlation functions which
appear in these equations are of the form 
\begin{equation}
\left(\mathrm{polynomial}\,\mathrm{of}\, M\right)
 \times e^{MQ_{4}}\,,
\label{eq:polynomialeMQ}
\end{equation}
 as functions of $M$, where $Q_{4}$ is given in eq.(\ref{eq:def-QN})
with $N=4$.
These correlation functions consist of the
contributions from the $X^{\pm}$ CFT, 
the transverse part, the super-reparametrization ghost part 
and the superconformal field theory in 
section~\ref{sec:Dimensional-regularization}.
$M$ dependence comes either from the $X^{\pm}$ CFT 
or the superconformal field theory. 
With the external states~(\ref{eq:external}), the
contributions from the latter are the combinatorial factors again
and can be expressed as a polynomial of $M$. Therefore what we should
show is that the contributions from the $X^{\pm}$ CFT are of the
form (\ref{eq:polynomialeMQ}).

The $M$ dependence of the contributions from the $X^{\pm}$ CFT can
be seen as follows. In these correlation functions, the variables
$X^{\pm},\psi^{\pm},\tilde{\psi}^{\pm}$ appear in 
\begin{itemize}
\item the picture changing operators $X,\tilde{X}$, 
\item $\mathcal{S}_{r}^{-1}$ 
\item the vertex operators $V_{r}^{\left(p_{L,r},p_{R,r}\right)}$. 
\end{itemize}
$V_{r}^{\left(p_{L,r},p_{R,r}\right)}$ involves a factor 
\begin{equation}
e^{-ip_{r}^{+}X^{-}
   -i\left(p_{r}^{-}-\frac{\mathcal{N}}{p_{r}^{+}}
   +\frac{d-10}{16}\frac{1}{p_{r}^{+}}\right)X^{+}}
 \left(Z_{r},\bar{Z}_{r}\right)
= e^{-ip_{r}^{+}X^{-}
     -\frac{i}{2p_{r}^{+}}\left(\vec{p}_{r}^{2}-1\right)X^{+}}
 \left(Z_{r},\bar{Z}_{r}\right)\,,
\end{equation}
 and the correlation functions are given by the path integral 
of the form 
\begin{eqnarray}
 & & \int\left[dX^{\pm}d\psi^{\pm}d\tilde{\psi}^{\pm}\right]
      e^{-S_{X^{\pm}}}
      \prod_{r}e^{-ip_{r}^{+}X^{-}}\left(Z_{r},\bar{Z}_{r}\right)
      \cdots
\nonumber \\
 & & \quad=\int\left[dX^{\pm}d\psi^{\pm}d\tilde{\psi}^{\pm}\right]
      e^{\frac{1}{\pi}\int d^{2}z
          \left(\partial X^{+}\bar{\partial}X^{-}
                 +\psi^{+}\bar{\partial}\psi^{-}
                 +\tilde{\psi}^{+}\partial\tilde{\psi}^{-}
           \right)}
\nonumber \\
 &  & 
     \hphantom{\quad=\int\left[dX^{\pm}
               d\psi^{\pm}d\tilde{\psi}^{\pm}\right]}
    \times e^{-\frac{d-10}{8}\Gamma_{\mathrm{super}}
                  \left[X^{+},\psi^{+},\tilde{\psi}^{+}\right]
              -\sum_{r}ip_{r}^{+}X^{-} 
                         \left(Z_{r},\bar{Z}_{r}\right)
              }
     \, \cdots\,,
\label{eq:pathintegral}
\end{eqnarray}
 where the ellipses denote the other insertions,
$S_{X^{\pm}}$ is the action of the $X^{\pm}$ CFT
given in Ref.~\cite{Baba:2009fi}
and
\begin{equation}
\Gamma_{\mathrm{super}}\left[X^{+},\psi^{+},\tilde{\psi}^{+}\right]
  \equiv  -\frac{1}{2\pi}\int d^{2}\mathbf{z}\bar{D}\Phi D\Phi\,.
\end{equation}
With the insertion 
$\prod_re^{-\sum_{r}ip_{r}^{+}X^{-}} \left(Z_{r},\bar{Z}_{r}\right)$, 
the variable $X^+$ possesses an expectation value. 
Dividing the variable
$X^{+}$ and the functional $\Gamma_{\mathrm{super}}$
into their
expectation values and fluctuations as 
\begin{eqnarray}
X^{+} 
 &=& -\frac{i}{2}\left(\rho\left(z\right)
                       +\bar{\rho}\left(\bar{z}\right)\right)
     +\delta X^{+}\,,
\nonumber \\
\Gamma_{\mathrm{super}}\left[X^{+},\psi^{+},\tilde{\psi}^{+}\right]
 &=& \left\langle \Gamma_{\mathrm{super}}\right\rangle 
     +\delta\Gamma_{\mathrm{super}}\,,
\nonumber \\
\left\langle \Gamma_{\mathrm{super}}\right\rangle  
 & \equiv & \Gamma_{\mathrm{super}}
             \left[-\frac{i}{2}\left(\rho\left(z\right)
                            +\bar{\rho}\left(\bar{z}\right)\right)
                   ,0,0\right]\,,
\end{eqnarray}
eq.(\ref{eq:pathintegral}) becomes 
\begin{eqnarray}
 &  & \int\left[d\delta X^{+} dX^{-}
               d\psi^{\pm}d\tilde{\psi}^{\pm}\right]
   e^{\frac{1}{\pi}\int d^{2}z\left(
           \partial\delta X^{+}\bar{\partial}X^{-}
            +\psi^{+}\bar{\partial}\psi^{-}
            +\tilde{\psi}^{+}\partial\tilde{\psi}^{-}\right)}
\nonumber \\
 &  & \hphantom{\quad\int\left[d\delta X^{+}dX^{-}
                d\psi^{\pm}d\tilde{\psi}^{\pm}\right]}
 \times e^{-\frac{d-10}{8}\left\langle \Gamma_{\mathrm{super}}
                          \right\rangle }
      \sum_{n=0}^{\infty}
       \frac{\left(-\frac{d-10}{8}\delta\Gamma_{\mathrm{super}}
             \right)^{n}}{n!}
       \cdots\,.
\label{eq:perturabative}
\end{eqnarray}
$\left\langle \Gamma_{\mathrm{super}}\right\rangle $ coincides with
$\frac{1}{2}\Gamma$. 
We can calculate the correlation functions perturbatively 
with $d-10$ as the coupling constant, 
by contracting 
$\left(\delta\Gamma_{\mathrm{super}}\right)^{n}$ and
$\cdots$ by using the free propagators for
$\delta X^{+},X^{-},\psi^{\pm},\tilde{\psi}^{\pm}$.%
\footnote{The correlation functions in the $X^{\pm}$ CFT 
was defined and calculated in Ref.~\cite{Baba:2009fi}. 
Since the perturbative expansion here terminates
at a finite order and one can obtain exact results, they coincide
with those given there. 
It is also possible to interpret
the definition in Ref.~\cite{Baba:2009fi} 
directly in terms of the perturbative
calculations. %
} 

It is convenient to introduce an additive quantum number which we
refer to as the $\pm$ number assigned to the fields 
as \cite{Baba:2009zm}
$$
\begin{array}{c||c|c|c|c|c}
\mathrm{fields} & \delta X^{+} & X^{-} & \psi^{\pm} & \tilde{\psi}^{\pm} 
& \mathrm{others}\\
\hline \pm\mathrm{number} & +1 & -1 & \pm1 & \pm1 & 0
\end{array}\,.
$$
In eq.(\ref{eq:perturabative}) only the terms in 
$\sum_{n=0}^{\infty}
  \frac{\left(-\frac{d-10}{8}\delta\Gamma\right)^{n}}{n!}\cdots$
with vanishing $\pm$ number have nonvanishing contributions. 

In the perturbative calculation, 
the contribution of each ingredient given
above is summarized as 
\begin{itemize}
\item $X,\tilde{X}$ contribute factors of 
      $\partial\delta X^{+},\bar{\partial}\delta X^{+},\partial X^{-},
      \bar{\partial}X^{-},\psi^{\pm},\tilde{\psi}^{\pm}$. 
\item $\mathcal{S}_{r}^{-1}$ has the expectation value 
      $e^{\frac{d-10}{16}\frac{\tau_{0}^{\left(r\right)}}{p_{r}^{+}}}$
       and the contribution is of the form 
\begin{equation}
e^{\frac{d-10}{16}\frac{\tau_{0}^{\left(r\right)}}{p_{r}^{+}}}
 \left( 1 + \sum_{l=1}^{\infty}
              \left(d-10\right)^{l}
              \mathcal{O}_{l}
                 \left( z_{I^{\left(r\right)}},
                        \bar{z}_{I^{\left(r\right)}}\right)\right)\,,
\end{equation}
 where $\mathcal{O}_{l}$ denotes a local operator 
  made from derivatives of 
  $\delta X^{+},X^{-},\psi^{\pm},\tilde{\psi}^{\pm}$ 
  with the $\pm$ number larger than or equal to $l$. 
\item $V_{r}^{\left(p_{L,r},p_{R,r}\right)}$ includes spin fields. 
      They are defined by using the free field description given in 
      Ref.~\cite{Ishibashi:2010nq}.
      They can be rewritten in terms of 
      $\delta X^{+},X^{-},\psi^{\pm},\tilde{\psi}^{\pm}$,
      but the expression is quite complicated. 
      However, here we do not need
      the exact form of the operator. 
      The $-\frac{3}{2}$ picture vertex operator involves a spin field 
      of the form 
\begin{equation}
e^{\sigma-\frac{3}{2}\phi}e^{\pm\frac{i}{2}H}
 \left(1+\sum_{l=1}^{\infty}\left(d-10\right)^{l}
                 \mathcal{O}_{l}^{\prime}
                    \left(Z_{r},\bar{Z}_{r}\right)
\right)\,,
\end{equation}
  where $i\partial H=\psi^{-}\psi^{+}$, 
  which can be used for bosonization in the perturbative treatment. 
  $\mathcal{O}_{l}^{\prime}$ is a local operator made from 
  derivatives of $\delta X^{+},X^{-},\psi^{\pm},\tilde{\psi}^{\pm}$
  with the $\pm$ number larger than or equal to $l$. 
  The vertex operators in other pictures in the Ramond sector 
  are expressed by acting the picture changing operators on 
  the $-\frac{3}{2}$ picture ones. 
  The picture changing operators contribute factors of 
     $\partial\delta X^{+},\bar{\partial}\delta X^{+},
      \partial X^{-},\bar{\partial}X^{-},\psi^{\pm},\tilde{\psi}^{\pm}$,
  as stated above.
\end{itemize}
Altogether, the correlation functions are given as sums of the terms
\begin{eqnarray}
 && e^{-\frac{d-10}{8}
         \left\langle \Gamma_{\mathrm{super}}\right\rangle
       +\sum_{r}\frac{d-10}{16}
         \frac{\tau_{0}^{\left(r\right)}}{p_{r}^{+}}}
    \left(d-10\right)^{n+l}
\nonumber \\
 &  & \quad\times\int\left[d\delta X^{+} dX^{-}
                           d\psi^{\pm}d\tilde{\psi}^{\pm}\right]
      e^{\frac{1}{\pi}\int d^{2}z
         \left(\partial\delta X^{+}\bar{\partial}X^{-}
                +\psi^{+}\bar{\partial}\psi^{-}
                +\tilde{\psi}^{+}\partial\tilde{\psi}^{-}\right)}
\nonumber \\
 &  & \hphantom{\quad\times\int\left[d\delta X^{+}dX^{-}
                d\psi^{\pm}d\tilde{\psi}^{\pm}\right]}
   \times \left(\delta\Gamma_{\mathrm{super}}\right)^{n}
          \mathcal{O}_{l}
\nonumber \\
 &  & \hphantom{\quad\times\int\left[d\delta X^{+}dX^{-}
                d\psi^{\pm}d\tilde{\psi}^{\pm}\right]}
\times \partial X^{-} \cdots
       \bar{\partial}X^{-} \cdots
       \psi^{-} \cdots
       \tilde{\psi}^{-} \cdots \,.
\end{eqnarray}
The factors 
 $\partial X^{-}\cdots\bar{\partial}X^{-}\cdots
  \psi^{-}\cdots\tilde{\psi}^{-}\cdots$
come from the picture changing operators,
and the number of these factors is less than that of 
the picture changing operators in the correlation functions. 
In our case, there are five picture changing operators including
the one to express 
$V_{2}^{\left(-\frac{1}{2},-1\right)}(Z_{2},\bar{Z}_{2})$
as 
\begin{equation}
V_{2}^{\left(-\frac{1}{2},-1\right)}(Z_{2},\bar{Z}_{2})
 =XV_{2}^{\left(-\frac{3}{2},-1\right)}(Z_{2},\bar{Z}_{2})\,.
\end{equation}
Since only the terms with vanishing $\pm$ number can
contribute,
the perturbative expansion terminates 
at a finite order. 
Using
\begin{equation}
d-10 = -\frac{8M}{k+2}\,,
\qquad
-\frac{d-10}{8}\left\langle \Gamma_{\mathrm{super}}\right\rangle
+\sum_{r=1}^{4}\frac{d-10}{16}\frac{\tau_{0}^{\left(r\right)}}{p_{r}^{+}} 
 =  MQ_{4}\,,
\end{equation}
we can see that the correlation functions 
are of the form (\ref{eq:polynomialeMQ}).
%%can be written
%%as a polynomial of $M$ times $e^{MQ_{4}}$. 
It is easy to generalize the arguments here to more general correlation 
functions, such as those which appear in $N$ point amplitudes. 

Thus we have proved that all the correlation functions which appear
on the right hand side of eq.(\ref{eq:A42}) are expressed as 
eq.(\ref{eq:polynomialeMQ}).
Therefore we define them as analytic functions of $M$ 
using this expression.
We define 
eq.(\ref{eq:A4finite}) for noninteger $M$ 
in the same way.

The integrand on the right hand side of eq.(\ref{eq:A42}) can be
written in the form 
\begin{equation}
R_{4}\left(M\right)e^{MQ_{4}}
+\partial_{\mathcal{T}}\left(S_{4}\left(M\right)e^{MQ_{4}}\right)
+\partial_{\bar{\mathcal{T}}}\left(\bar{S}_{4}\left(M\right)
                                    e^{MQ_{4}}\right)\,,
\label{eq:expression}
\end{equation}
where 
$R_{4}\left(M\right),S_{4}\left(M\right),
  \bar{S}_{4}\left(M\right)$
are polynomials of $M$. 
Here we make only the $M$ dependence explicit
but the coefficients of these polynomials of course depend on 
$\mathcal{T},\bar{\mathcal{T}}$ and other kinematic parameters. 
Since the integrand on the right hand side of eq.(\ref{eq:A4}) 
is written as $P_{4}\left(M\right)e^{MQ_{4}}$,
the relation 
\begin{equation}
P_{4}\left(M\right)e^{MQ_{4}}
 = R_{4}\left(M\right)e^{MQ_{4}}
   + \partial_{\mathcal{T}}\left(S_{4}\left(M\right)
                                 e^{MQ_{4}}\right)
   + \partial_{\bar{\mathcal{T}}}
           \left(\bar{S}_{4}\left(M\right)
                 e^{MQ_{4}}\right)
\label{eq:identity4}
\end{equation}
 holds if $M$ is a non-negative integer. It implies that 
\begin{equation}
P_{4}\left(M\right)
= R_{4}\left(M\right)
  + \partial_{\mathcal{T}} S_{4}\left(M\right)
  + S_{4}\left(M\right)M \partial_{\mathcal{T}}Q_{4}
  + \partial_{\bar{\mathcal{T}}} \bar{S}_{4}\left(M\right)
  + \bar{S}_{4}\left(M\right)M \partial_{\bar{\mathcal{T}}}Q_{4}\,.
\label{eq:polynomial}
\end{equation}
Since 
 $P_{4}\left(M\right),R_{4}\left(M\right),
  S_{4}\left(M\right),\bar{S}_{4}\left(M\right)$
are all polynomials,
eq.(\ref{eq:polynomial}) can be written as 
\begin{equation}
\sum_{k=0}^{n}c_{k}M^{k}=0\,.
\end{equation}
 If such an identity holds for any non-negative integer $M$, then
$c_{k}=0$ and thus it holds for any complex number $M$. Therefore
eq.(\ref{eq:A42}) holds for the analytically continued correlation functions.

Now that the right hand side of eq.(\ref{eq:A42})
is defined as an analytic function of $M$,
discarding the total derivative terms, we obtain  
\begin{equation}
\mathcal{A}_{4}
 \sim \int d^{2}\mathcal{T}
      \left\langle 
            \prod_{A}\left(c^{A}\left(z_{0}\right)
                           \tilde{c}^{A}\left(\bar{z}_{0}\right)
                            \right)
            \left|\oint_{C}\frac{dz}{2\pi i}
                  \frac{b}{\partial\rho}\left(z\right)
            \right|^{2}
            \prod_{r=1}^{4}
               \left[\mathcal{S}_{r}^{-1}
                     V_{r}^{\left(p_{L,r}^{\prime},
                                  p_{R,r}^{\prime}\right)}
               \right]
      \right\rangle \,,
\end{equation}
as an analytic function of $M$.
 In the limit $M\to0$, $\mathcal{A}_{4}$ coincides with the result
of the first quantized formalism.

\subsubsection*{General case}

It is straightforward to generalize the above arguments to the case
of $N$ point amplitudes. 
The conformal gauge expression
for the $N$ point amplitudes (\ref{eq:AN-1}) with eq.(\ref{eq:LCFN})
becomes 
\begin{eqnarray}
\mathcal{A}_{N} 
 & \sim & 
   \int\prod_{\mathcal{I}=1}^{N-3} d^{2}\mathcal{T}_{\mathcal{I}}
   \left\langle \prod_{A}\left(c^{A}\left(z_{0}\right)
                               \tilde{c}^{A}\left(\bar{z}_{0}\right)
                         \right)
  \right.   \nonumber \\
& &
   \hphantom{\int\prod_{\mathcal{I}}d^{2}\mathcal{T}_{\mathcal{I}}\quad}
   \left. \times
    \prod_{\mathcal{I}=1}^{N-3}
     \left| \oint_{C_{\mathcal{I}}}\frac{dz}{2\pi i}
              \frac{b}{\partial\rho} 
     \right|^{2}
    \prod_{I=1}^{N-2} \left|X\left(z_{I}\right)\right|^{2}
    \prod_{r=1}^{N}
     \left[\mathcal{S}_{r}^{-1}V_{r}^{\left(p_{L,r},p_{R,r}\right)}
     \right]
   \right\rangle,
\label{eq:FNpic}
\end{eqnarray}
where 
%%$\mathcal{T}_{\mathcal{I}},\bar{\mathcal{T}}_{\mathcal{I}}$
%%are the moduli parameters, which correspond to 
%%the Schwinger parameters
%%for the propagators in each channel and 
the integration contour $C_{\mathcal{I}}$
is taken to go around the $\mathcal{I}$-th internal propagator. 
%%$\rho\left(z\right)$
%%is the Mandelstam mapping in eq.(\ref{eq:Mandelstam}) 
%%and $z_{I}$ are the interaction points. 
The BRST invariance of this expression
can be shown by using 
\begin{equation}
\left\{ Q_{\mathrm{B}},
        \oint_{C_{\mathcal{I}}}\frac{dz}{2\pi i}
              \frac{b}{\partial\rho}
\right\} 
\sim -\partial_{\mathcal{T}_{\mathcal{I}}}\,.
\end{equation}

One can show that the right hand side of eq.(\ref{eq:FNpic}) 
is equal to 
\begin{equation}
\int\prod_{\mathcal{I}=1}^{N-3}
 d^{2}\mathcal{T}_{\mathcal{I}}
 \left\langle 
    \prod_{A}\left(c^{A}\left(z_{0}\right)
                   \tilde{c}^{A}\left(\bar{z}_{0}\right)\right)
    \prod_{\mathcal{I}=1}^{N-3}
        \left|\oint_{C_{\mathcal{I}}}\frac{dz}{2\pi i}
                \frac{b}{\partial\rho}
        \right|^{2}
    \prod_{r=1}^{N}\left[\mathcal{S}_{r}^{-1}
                         V_{r}^{\left(p_{L,r}^{\prime},
                                      p_{R,r}^{\prime}\right)}
                   \right]
  \right\rangle \ ,
\label{eq:finite}
\end{equation}
up to total derivatives with respect to 
$\mathcal{T}_{\mathcal{I}},\bar{\mathcal{T}}_{\mathcal{I}}$,
by moving the picture changing operators $X(z_{I}),\tilde{X}(\bar{z}_{I})$
to the positions of external lines. Since eq.(\ref{eq:finite}) coincides
with the results of the first quantized formalism for $M=0$, 
what we should show is 
that $\mathcal{A}_{N}$ is equal to (\ref{eq:finite}) even for
noninteger $M.$

The correlation functions which appear in 
eqs.(\ref{eq:FNpic}) and (\ref{eq:finite})
are respectively expressed as 
$P_{N}\left(M\right) e^{MQ_{N}}$ and
$R_{N}\left(M\right)e^{MQ_{N}}$
using polynomials $P_{N}\left(M\right)$ and $R_{N}\left(M\right)$, 
as we have mentioned above. 
We define them as analytic functions of $M$ using these expressions.
For $M$ non-negative integer, 
by moving the picture changing operators, we can see that
\begin{equation}
\left(P_{N}\left(M\right)-R_{N}\left(M\right)\right)e^{MQ_{N}}
\end{equation}
 is expressed as derivatives of correlation functions with respect
to the moduli parameters. 
Those correlation functions are also written
in the form
\begin{equation}
\left(\mathrm{polynomial}\,\mathrm{of}\, M\right)
 \times e^{MQ_{N}}\,.
\end{equation}
Thus, when $M$ is
a non-negative integer, there exist the polynomials 
$S_{N}^{\mathcal{I}}(M),\bar{S}_{N}^{\mathcal{I}}(M)$
of $M$ such that 
\begin{equation}
\left( P_{N}\left(M\right) - R_{N}\left(M\right) \right) e^{MQ_{N}}
= \sum_{\mathcal{I}=1}^{N-3}
   \left[ \partial_{\mathcal{T}_{\mathcal{I}}}
            \left(S_{N}^{\mathcal{I}}\left(M\right)e^{MQ_{N}}\right)
         + \partial_{\bar{\mathcal{T}}_{\mathcal{I}}}
            \left(\bar{S}_{N}^{\mathcal{I}}\left(M\right)e^{MQ_{N}}
            \right)
   \right]\ .
\label{eq:identity}
\end{equation}
%where $P_{\mathcal{I}}\left(M\right)$, 
%$\bar{P}_{\mathcal{I}}\left(M\right)$ are polynomials
%of $M$, when $M$ is a positive integer. 
Eq.(\ref{eq:identity}) 
implies that a polynomial equation 
\begin{equation}
P_{N} (M) - R_{N} (M) 
= \sum_{\mathcal{I}=1}^{N-3}
   \left[ 
   \partial_{\mathcal{T}_{\mathcal{I}}}S_{N}^{\mathcal{I}}
 + S_{N}^{\mathcal{I}}
    M \partial_{\mathcal{T}_{\mathcal{I}}}Q_{N}
 + \partial_{\bar{\mathcal{T}}_{\mathcal{I}}}
    \bar{S}_{N}^{\mathcal{I}} 
 + \bar{S}_{N}^{\mathcal{I}}
    M \partial_{\bar{\mathcal{T}}_{\mathcal{I}}}Q_{N}
   \right]
\end{equation}
is satisfied for all non-negative integers $M$, and we can conclude that  
 it holds for any complex number $M$. 
Therefore the analytically continued amplitudes
are given as eq.(\ref{eq:finite}). 
Thus the dimensionally regularized
amplitudes are finite and become equal to the first quantized results
as $M\to0$.

\section{Conclusions and discussions 
  \label{sec:Conclusions-and-discussions}}

In this paper, we have shown that the previously proposed dimensional
regularization 
scheme~\cite{Baba:2009kr,Baba:2009ns,Baba:2009fi,Baba:2009zm}
also works for the tree-level amplitudes involving the external lines
in the Ramond sector in the light-cone gauge 
type~II superstring field theories.
In order to overcome the difficulty pointed out 
in Ref.~\cite{Ishibashi:2010nq},
we have modified the worldsheet theory. We have shown that the tree-level
amplitudes calculated using the modified worldsheet theory can be
analytically continued and used to regularize the superstring field
theories. The resultant amplitudes turn out to coincide with those
of the first quantized formulation without introducing any contact
terms as counterterms.

\begin{figure}[h]
\begin{centering}
\includegraphics[scale=0.5]{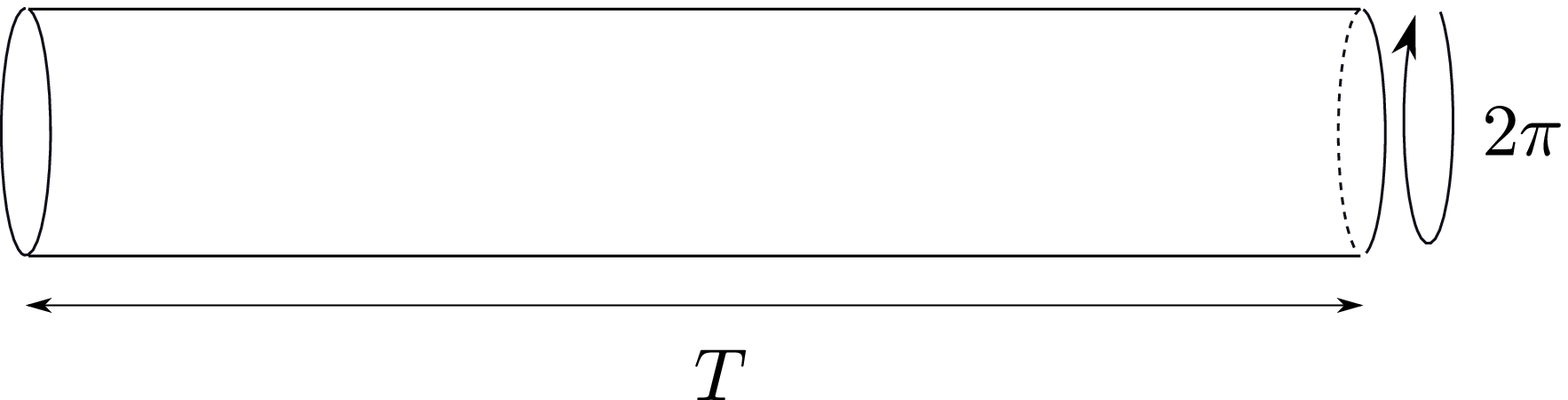} 
\par\end{centering}

\caption{Tube\label{fig:tube}}

\end{figure}

Therefore, with the regularization, 
the light-cone gauge closed superstring field
theory with only cubic interaction terms correctly describes the tree-level
amplitudes. 
One obvious thing to be done is to examine if our dimensional
regularization scheme can be employed to regularize the ultraviolet
divergences of the multi-loop amplitudes. One can argue that the dimensional
regularization can be used to regularize the divergences of the multi-loop
amplitudes as in the field theory for particles, 
although the way how the regularization works in string field theory 
is a little bit different.
%% from that in 
%%the field theory for particles. 
In string theory, ultraviolet
divergences in some channel correspond to  infrared divergences
in another channel. Therefore what we should do is to regularize the
infrared divergences. Infrared divergences come from the long tube-like
worldsheet depicted in Fig.~\ref{fig:tube}. If the worldsheet theory
is taken to be a CFT with central charge $c$, 
the contribution of a worldsheet
which includes such a cylinder behaves as 
\begin{equation}
\sim\exp\left(T\left(\frac{c}{12}-2\bigtriangleup_{min}\right)\right)\ ,
\end{equation}
for $T\gg1$. 
 Here $\bigtriangleup_{min}$ denotes the conformal dimension of the
primary field with the lowest dimension in the worldsheet CFT. Therefore,
if one can use a CFT with $\frac{c}{12}-2\bigtriangleup_{min}$ largely
negative in the dimensional regularization, we expect that the divergences
are regularized. In the case of the superconformal field theory proposed
in section~\ref{sec:Dimensional-regularization}, 
\begin{eqnarray}
\frac{c}{12}-2\bigtriangleup_{min} 
 & = & \frac{c}{12} 
  \nonumber \\
 & = & 1-\frac{M}{k+2}\ ,
\end{eqnarray}
 and one can make the amplitudes finite by taking $M$ large enough.
Besides such a stringy aspect, 
our dimensional regularization scheme has a lot in common with the 
dimensional reduction scheme employed in the field theory for particles. 
Therefore the regularization scheme we propose in this paper will
be useful to regularize the ultraviolet divergences of the multi-loop
amplitudes. 
It will be intriguing to compare the amplitudes defined by using 
our regularization with those given in Ref.~\cite{D'Hoker:2002gw}.

One important problem to be addressed about the multi-loop calculations 
is the definition 
of $(-1)^F$. 
In the field theory for particles, 
the dimensional regularization has problems
in treating $\gamma^5$. 
In order to keep 
$\mathrm{tr}\gamma^{5}
            \gamma^{\mu}\gamma^{\nu}\gamma^{\rho}\gamma^{\sigma}$
nonvanishing for general $d$, the definition of $\gamma^{5}$ should
be taken so that it does not anticommute with 
$\gamma^{\mu}$~\cite{'tHooft:1972fi,Akyeampong:1973xi,Breitenlohner:1977hr}.
In superstring theory, this problem is related to the definition of
$\left(-1\right)^{F}$ which appears in the definition of the GSO
projection operator for the Ramond sector. $\left(-1\right)^{F}$
should anticommute with $\psi^{\mu}$ for the BRST invariance and
it seems that we will have trouble in defining it for general $d$.
Indeed, the superconformal field theory defined 
in section~\ref{sec:Dimensional-regularization}
is not well-defined for odd spin structure. The partition function
of the $\beta\gamma$ system is the inverse of that of the $\psi$
system and therefore it is divergent. The odd spin structure arises
when we insert $\left(-1\right)^{F}$ and the problem is related to
the definition of this operator. One way to deal with the divergence
is to modify the definition of $\left(-1\right)^{F}$. Doing so breaks
the gauge invariance but it may be possible to prove that the procedure
does not cause any problems if the theory is not anomalous. 

Whatever the outcome of the multi-loop calculations will be, 
the cubic light-cone gauge action is valid classically. 
Such a formulation may be useful in exploring classical solutions 
of closed superstring field theories. 
For many reasons, 
it would be better to have a gauge invariant version of such a theory. 
Since we have written down the conformal gauge expression 
of the amplitudes, 
we can infer what the gauge invariant version would be. 
For the bosonic string theory, 
as we have pointed out in Ref.~\cite{Baba:2009ns}, 
it should be the $\alpha =p^+$ HIKKO theory~\cite{Kugo:1992md}. 
Construction of such a theory for superstrings will be a problem 
to be pursued. 

We would like to come back to these problems elsewhere.

%%%
\section*{Acknowledgements}

We would like to thank the Yukawa Institute for Theoretical Physics 
at Kyoto University, 
where some of the results of this work were presented at 
the YITP-W-10-13 on ``String Field Theory and Related Aspects",
which was supported  by the Grant-in-Aid for the Global COE Program
``The Next Generation of Physics, Spun from Universality and
Emergence" from the Ministry of Education, Culture, Sports, Science and
Technology (MEXT) of Japan.
One of us (N.I.) would like to acknowledge the hospitality of 
Okayama Institute for Quantum Physics
where part of this work was done.
This work was supported in part by 
Grant-in-Aid for Scientific Research~(C) (20540247) from MEXT.
%%%%%%%%%%%%%%%%%%%%%%%%%%%%%%%%%%%%%%%%%%%%%%%%%%%%%%%%%%%%%%%

\appendix

\section{Action and amplitudes for $d\neq10$}

\label{sec:SFTaction}

In this appendix, we present the action of the light-cone gauge superstring
field theory for $d\neq10$, and calculate the tree-level amplitudes
perturbatively by using this action. 
%The contributions of the (NS,NS)
%and the (R,R) sectors in the naive dimensional regularization have been 
%calculated in Refs.~\cite{Baba:2009zm,Ishibashi:2010nq}.

The string fields 
are taken to be GSO even and satisfy 
the level-matching condition.
Those in the bosonic sector, namely the (NS,NS) or the (R,R) sector,
are Grassmann even, whereas those in the fermionic sector,
namely the (R,NS) or the (NS,R) sector, are Grassmann odd.

\subsection{The kinetic term and the propagator}

The kinetic terms and the propagators for the string fields in the
bosonic sector are given 
in Refs.~\cite{Baba:2009zm,Ishibashi:2010nq}.\footnote{%%%
     Although the definitions in those references are 
     given for the naive dimensional regularization, 
     it is straightforward to generalize them 
      to any worldsheet theory as will be mentioned 
      at the end of this subsection.}
For the string fields in the fermionic sector, 
%%namely those in the
%%(R,NS) and the (NS,R) sectors, 
the kinetic term is given as 
\begin{equation}
\frac{1}{2}\int dt\int d1d2
   \left\langle R\left(1,2\right)|\Phi(t)\right\rangle _{1}
   \left(i\frac{\partial}{\partial t}
         -\frac{L_{0}^{\mathrm{LC}(2)}+\tilde{L}_{0}^{\mathrm{LC}(2)}
                 -\frac{d-2}{8}}
               {\alpha_{2}}
    \right)
    \left|\Phi(t)\right\rangle _{2}\ .
\label{eq:kinetic}
\end{equation}
 Here $t=x^{+}$, $dr$ and $\alpha_{r}=2p_{r}^{+}$ are the zero-mode
measure and the string-length parameter for the $r$-th string respectively,
and $\left\langle R(1,2)\right|$ is the reflector. $L_{0}^{\mathrm{LC}(r)}$
denotes the Virasoro zero mode of the light-cone gauge worldsheet
theory. While the kinetic term~(\ref{eq:kinetic}) is in the same
form as that for the strings in the bosonic sector, this time we should
differently define  the zero-mode measure and the reflector.
The zero-mode measure is defined as
\begin{equation}
dr = 
   \frac{d\alpha_{r}}{4\pi}\frac{d^{d-2}p_{r}}{\left(2\pi\right)^{d-2}}
   =\frac{dp_{r}^{+}}{2\pi}\frac{d^{d-2}p_{r}}{\left(2\pi\right)^{d-2}}\ .
\end{equation}
Compared with $dr$ for the bosonic sector, a factor of $\alpha_{r}$
is absent in this case.
The reflector is defined as
\begin{equation}
\left\langle R\left(1,2\right)\right| 
   = 
   \delta\left(1,2\right)
    %%%{}_{12}\left\langle 0\right|e^{E\left(1,2\right)}~,
    \sum_{n} {}_{2}\langle n| {}_{1} \langle n|~,
\end{equation}
where
\begin{equation}
\delta\left(1,2\right) 
    =  
     4\pi\delta\left(\alpha_{1}+\alpha_{2}\right)
     \left(2\pi\right)^{d-2}\delta^{d-2} \left(p_{1}+p_{2}\right)~,
\end{equation}
and we have introduced
a basis $\left\{ \left|n\right\rangle \right\} $
of the projected Fock space for the non-zero modes
and the fermionic zero modes.
$|n\rangle$ is Grassmann even and normalized as
\begin{equation}
\left\langle n|n'\right\rangle
    =\delta_{n,n'}~.
\end{equation}

As in the bosonic case, by using the basis $\{|n\rangle\}$,
the string field 
$\left|\Phi\right\rangle $ can be expanded as 
\begin{equation}
\left|\Phi(t)\right\rangle 
=\sum_{n}\psi_{n}\left(t,\alpha,\vec{p}\right)\left|n\right\rangle ,
\end{equation}
 and 
\begin{equation}
\left(L_{0}^{\mathrm{LC}}+\tilde{L}_{0}^{\mathrm{LC}}-\frac{d-2}{8}\right)
  \left|\Phi(t)\right\rangle 
  = \sum_{n} \left(\vec{p}^{\;2}+m_{n}^{2}\right)
      \psi_{n}\left(t,\alpha,\vec{p}\right)\left|n\right\rangle .
\label{eq:fermionic-modes}
\end{equation}
 This time, $\psi_{n}$ is Grassmann odd. 
In terms of $\psi_{n}$, the kinetic term (\ref{eq:kinetic})
is written as 
\begin{equation}
 \frac{1}{2}\sum_{n}\int\frac{d^{d}p}{\left(2\pi\right)^{d}}
    \tilde{\psi}_{n}\left(-p\right)
    \left(p^{-}-\frac{\vec{p}^{2}+m_{n}^{2}}{2p^{+}}\right)
    \tilde{\psi}_{n}\left(p\right)\ ,
\label{eq:kineticpsi}
\end{equation}
 where
\begin{equation}
\tilde{\psi}_{n}(p)\equiv\int dt\, e^{ip^{-}t}\psi_{n}(t,\alpha,\vec{p})~.
\end{equation}
 In order for the kinetic term (\ref{eq:kineticpsi}) to be consistent
with the fact that $\psi_{n}$ is Grassmann odd, we have chosen $dr$
different from that for the bosonic case. With the kinetic term thus
defined the propagator becomes
\begin{eqnarray}
\begC2{\tilde{\psi}}
  \conC{_{n}\left(p\right)}
  \endC2{\tilde{\psi}}_{n^{\prime}}\left(p'\right) 
 & = & \alpha\delta_{n,n'}
   \left(2\pi\right)^{d}\delta^{d}\left(p+p'\right)
   \frac{-i}{p^{2}+m_{n}^{2}}~,
 \nonumber \\
\Bigl|\begC2{\tilde{\Phi}}
      \conC{\left(p_{1}^{-}\right)  \Bigr\rangle_{1}
\Bigl|} \endC2{\tilde{\Phi}}\left(p_{2}^{-}\right)
  \Bigr\rangle_{2} 
 & = & \frac{1}{\alpha_{1}}\left(2\pi\right)^{d}
       \delta^{d}\left(p_{1}+p_{2}\right)
  \nonumber \\
 &  & 
    \hphantom{\frac{1}{\alpha_{1}}}
    \,\times\int\frac{d^{2}\mathcal{T}}{4\pi}\, 
        e^{-\frac{\mathcal{T}}{\left|\alpha_{1}\right|}
             \left(L_{0}^{\mathrm{LC}(1)}-\frac{d-2}{16}
             \right)
           -\frac{\bar{\mathcal{T}}}{\left|\alpha_{1}\right|}
              \left(\tilde{L}_{0}^{\mathrm{LC}(1)}-\frac{d-2}{16}
              \right)}
\nonumber \\
 &  & 
   \hphantom{\frac{1}{\alpha_{1}}}
    \,\times e^{\frac{\alpha_{1}}{\left|\alpha_{1}\right|}
                 p_{1}^{-}T}
      %%\mathcal{P}_{\mathrm{GSO}}^{(1)}
      %%\mathcal{P}_{\mathrm{GSO}}^{(2)}
      %%\, e^{E^{\dagger}\left(1,2\right)}
      %%\left|0\right\rangle _{12}~,
     \sum_{n} |n\rangle_{1} |n\rangle_{2}~,
\nonumber \\
\begC2{\tilde{\psi}} 
  \conC{_{n}\left(p\right)\Bigl|}
   \endC2{\tilde{\Phi}}\left(p'\right)\Bigr\rangle 
  & = & \alpha\left(2\pi\right)^{d}
        \delta^{d}\left(p+p'\right)
        \frac{-i}{p^{2}+m_{n}^{2}}\left|n\right\rangle ~.
\label{eq:PhiPhi2}
\end{eqnarray}
 Because of the different choice of $dr$, the propagators include
an extra factor of $\alpha$ compared with those for the bosons.

The kinetic terms and the propagators can be defined for any worldsheet
CFT. Here we would like to define the string field theory using the
worldsheet theory defined in section~\ref{sec:Dimensional-regularization}.
Since the ghost-like variables $b^{A},c^{A},\tilde{b}^{A},\tilde{c}^{A}$
have zero modes, we should mention how to treat them in defining the
string field theory. We take the string field $\left|\Phi\right\rangle $
to satisfy 
\begin{equation}
b_{0}^{A}\left|\Phi\right\rangle
  =\tilde{b}_{0}^{A}\left|\Phi\right\rangle =0\,.
\label{eq:b0A}
\end{equation}
 Then the propagators come with the projection operator corresponding
to the condition~(\ref{eq:b0A}).

\subsection{The three-string vertex\label{sub:The-three-string}}

The three-string vertex for the bosonic fields is given as 
\begin{eqnarray}
\lefteqn{
    \int dt\int d1d2d3\left\langle V_{3}\left(1,2,3\right)|
         \Phi(t)\right\rangle _{1}
         \left|\Phi(t)\right\rangle _{2}
         \left|\Phi(t)\right\rangle _{3}}
\nonumber \\
 &  & =\int\prod_{r=1}^{3}
         \left( \frac{d^{d}p_{r}}{\left(2\pi\right)^{d}}\alpha_{r}
         \right)
     \left(2\pi\right)^{d}
     \delta^{d}\Biggl(\sum_{r=1}^{3}p_{r}\Biggr)
     e^{-\Gamma^{\left[3\right]}\left(1,2,3\right)}
 \nonumber \\
 &  & 
   \hphantom{=\prod\int}
   \times\left\langle V_{3}^{\mathrm{LPP}}\left(1,2,3\right)\right|
      P_{123}
      \Bigl| \tilde{\Phi}\left(p_{1}^{-}\right) \Bigr\rangle_{1}
      \Bigl|\tilde{\Phi}\left(p_{2}^{-}\right)\Bigr\rangle_{2}
      \Bigl|\tilde{\Phi}\left(p_{3}^{-}\right)\Bigr\rangle_{3}~.
\label{eq:3vertex}
\end{eqnarray}
 Here $\left\langle V_{3}^{\mathrm{LPP}}\left(1,2,3\right)\right|$
denotes the LPP vertex~\cite{LeClair:1988sp,LeClair:1988sj}. 
$\Gamma^{(3)}(1,2,3)$
and $P_{123}$ are defined in eqs.(A.4) and (A.6) of 
Ref.~\cite{Baba:2009zm}.
With the vertex~(\ref{eq:3vertex}) and the propagator, one can calculate
the amplitudes~\cite{Baba:2009zm}. They are expressed in terms of
the correlation functions of vertex operators in the worldsheet theory.

%Even with spacetime fermions, we expect that the amplitudes can be
%given in a way similar to the bosonic case, but with different vertex
%operators. 
The propagators for the fermions come with extra factors
of $\alpha$. In order to make the theory Lorentz invariant in the
critical dimension, we should define the three-string vertex so as
to compensate for these extra factors of $\alpha$.
Thus the three-string vertex should be defined as
\begin{eqnarray}
\lefteqn{
  \int dt\int d1d2d3\left\langle V_{3}\left(1,2,3\right)|\Phi(t)
                    \right\rangle _{1}
         \left|\Phi(t)\right\rangle _{2}
         \left|\Phi(t)\right\rangle _{3}
}\nonumber \\
 &  & \propto
      \int\prod_{r=1}^{3}
    \left(\frac{d^{d}p_{r}}{\left(2\pi\right)^{d}}\alpha_{r}\right)
    \left(2\pi\right)^{d}\delta^{d}
      \Biggl(\sum_{r=1}^{3}p_{r}\Biggr)
    e^{-\Gamma^{\left[3\right]}\left(1,2,3\right)}
\nonumber \\
 &  & 
   \hphantom{=\prod\int}
   \times
     \left\langle V_{3}^{\mathrm{LPP}}\left(1,2,3\right)\right|
     P_{123}\alpha_{1}^{-\frac{1}{2}}
     \Bigl| \tilde{\Phi}\left(p_{1}^{-}\right) \Bigr\rangle_{1}
     \alpha_{2}^{-\frac{1}{2}}
     \Bigl| \tilde{\Phi}\left(p_{2}^{-}\right) \Bigr\rangle_{2}
     \Bigl| \tilde{\Phi}\left(p_{3}^{-}\right) \Bigr\rangle_{3}~,
\label{eq:3vertex2}
\end{eqnarray}
 for $\left|\Phi\right\rangle _{1},\left|\Phi\right\rangle _{2}$
fermionic and $\left|\Phi\right\rangle _{3}$ bosonic. Here the problem
is how to define the phase of $\alpha^{-\frac{1}{2}}$. Unless we
define the phase properly, the contributions to the amplitudes from
various channels do not connect smoothly and the theory will not become
Lorentz invariant when $d=10$.  
As far as we know, this problem has never been
discussed in the literature and we would like to discuss it in the
following.

\subsubsection*{Phase of $\alpha^{-\frac{1}{2}}$}

\begin{figure}[h]
\begin{centering}
\includegraphics[scale=0.5]{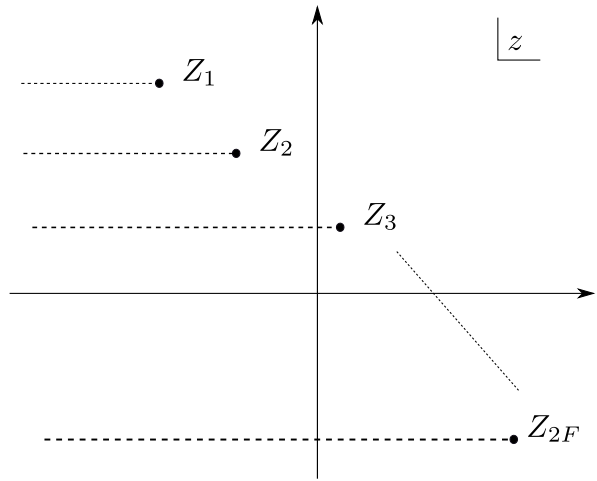} 
\par\end{centering}

\caption{Cuts are parallel to the real axis 
         and indicated by the dotted lines.\label{fig:cuts}}

\end{figure}

Suppose we perturbatively calculate 
the amplitude obtained by amputating
the external legs of
the following correlation function 
in the string field theory:
\begin{equation}
\left\langle \!\left\langle 
   \tilde{\psi}_{n_{1}}(p_{1})
   \,\tilde{\psi}_{n_{2}}(p_{2})
   \,\cdots  
    \,\tilde{\psi}_{n_{2F}}\left(p_{2F}\right)
    \,\tilde{\phi}_{n_{2F+1}}(p_{2F+1})
   \,\cdots\,
     \tilde{\phi}_{n_{N}}(p_{N})
\right\rangle \!\right\rangle ~.
\label{eq:SFTcorr}
\end{equation}
 Here $\langle\!\langle\cdots\rangle\!\rangle$ denotes the expectation
value in the string field theory. $\tilde{\phi}_{n}$ are the bosonic
modes of the string field given in Ref.~\cite{Ishibashi:2010nq},
whereas $\tilde{\psi}_{n}$ are the fermionic ones 
in eq.(\ref{eq:fermionic-modes}).
The correlation function (\ref{eq:SFTcorr}) can be evaluated by using
the propagators and the three-string vertices (\ref{eq:3vertex})
and (\ref{eq:3vertex2}). 
The amplitude this yields 
can be described as a sum of integrals
of the correlation function 
\begin{equation}
 \left\langle \prod_{I} 
          \left(T_{F}\left(z_{I}\right)
                \tilde{T}_{F}\left(\bar{z}_{I}\right)
          \right)
    \prod_{r=1}^{N}
%           \mathcal{O}_{n_{r}}\left(Z_{r},\bar{Z}_{r}\right)
V_r^\mathrm{LC}
\right\rangle _{\mathrm{LC}}\,\label{eq:correlationambi}
\end{equation}
 for the worldsheet theory on the $z$-plane, over the Schwinger parameters.
The vertex operator $V_r^\mathrm{LC}$ corresponds to the $r$-th external
line and is a local operator at $z=Z_r$. 
Due to the spin fields involved in 
$V_r^\mathrm{LC}$ 
and the supercurrent insertions at the interaction points $z_I$,
this correlation function can be defined only up to sign. Those
spin fields are expressed in the form $e^{iaH}\left(z\right)$, where
$H\left(z\right)$ is a chiral free boson and $a$ is a half integer.
Suppose we always define $e^{iaH} (z)$ by
\begin{equation}
e^{iaH}\left(z\right) 
 \equiv  \lim_{\Lambda\to\infty}\sum_{n=0}^{\infty}
       \frac{1}{n!}
       :\left(ia\int_{-\Lambda}^{0}dx\partial H\left(z+x\right)\right)^{n}:
 \ ,
\label{eq:define}
\end{equation}
 where the integration contour for $x$ is taken along the real axis.
Then the correlation function~(\ref{eq:correlationambi}) is defined
on the Riemann surface with cuts depicted in Fig.~\ref{fig:cuts}.

If one varies $Z_{r}$ continuously, the correlation function changes
its sign when some of $z_{I}$ cross the cuts. In order to remedy
this, we introduce 
\begin{eqnarray}
f\left(\alpha_{r};Z_{r}\right) 
  & \equiv & \prod_{n=1}^{F}\alpha_{2n}
    \prod_{I}\prod_{n=1}^{F}\left(z_{I}-Z_{2n-1}\right)^{\frac{1}{2}}
    \prod_{I}\prod_{n=1}^{F}\left(z_{I}-Z_{2n}\right)^{-\frac{1}{2}}
\nonumber \\
 &  & \quad\times
    \prod_{r=2F+1}^{N}\prod_{n=1}^{F}
          \left(Z_{r}-Z_{2n-1}\right)^{-\frac{1}{2}}
    \prod_{r=2F+1}^{N}\prod_{n=1}^{F}
          \left(Z_{r}-Z_{2n}\right)^{\frac{1}{2}}
\nonumber \\
 &  & \quad\times
      \prod_{F\ge n>n^{\prime}\ge1}
           \left(Z_{2n-1}-Z_{2n^{\prime}-1}\right)^{-1}
       \prod_{F\ge n>n^{\prime}\ge1}
           \left(Z_{2n}-Z_{2n^{\prime}}\right)\ ,
\label{eq:f}
\end{eqnarray}
 which coincides with
\begin{equation}
\prod_{r=1}^{2F}\left|\alpha_{r}\right|^{\frac{1}{2}}\ ,
\end{equation}
up to phase.
%\footnote{%%%
%   $f\left(\alpha_{r};Z_{r}\right)$ coincides with 
%   the ghost contributions 
%   to the covariant version of the amplitudes. 
%} 
Here we take the square-roots to be also defined with
the cuts depicted in Fig.~\ref{fig:cuts}. 
Then the value of the combination
\begin{equation}
f\left(\alpha_{r};Z_{r}\right)
 \left\langle 
    \prod_{I} \left(T_{F}\left(z_{I}\right)
                    \tilde{T}_{F}\left(\bar{z}_{I}\right)\right)
    \prod_{r=1}^{N}V_r^\mathrm{LC}
%          \mathcal{O}_{n_{r}}\left(Z_{r},\bar{Z}_{r}\right)
 \right\rangle _{\mathrm{LC}}
\label{eq:combination}
\end{equation}
 does not jump if we vary $Z_{r}$ continuously. This combination
can be proved to be invariant under $SL(2,\mathbb{C})$ up to sign.
Since it is not discontinuous, it is $SL(2,\mathbb{C})$ invariant.
It is easy to see that $f$ becomes $-f$ under the exchange 
$\left(\alpha_{2n-1},Z_{2n-1}\right)
   \leftrightarrow
   \left(\alpha_{2n^{\prime}-1}Z_{2n^{\prime}-1}\right)$
or 
$\left(\alpha_{2n},Z_{2n}\right)
   \leftrightarrow
   \left(\alpha_{2n^{\prime}},Z_{2n^{\prime}}\right)$
$\left(n\ne n^{\prime}\right)$.
Since
\begin{equation}
-\frac{\alpha_{2n-1}}{\alpha_{2n^{\prime}}} 
=  \frac{\prod_{I} 
             \left(z_{I}-Z_{2n-1}\right)
         \prod_{r\ne2n-1,2n^{\prime}}
             \left(Z_{r}-Z_{2n^{\prime}}\right)}
        {\prod_{I}\left(z_{I}-Z_{2n^{\prime}}\right)
         \prod_{r\ne2n-1,2n^{\prime}}\left(Z_{r}-Z_{2n-1}\right)}\ ,
\end{equation}
 one can prove that $f$ becomes $-f$ under the exchange 
$\left(\alpha_{2n-1},Z_{2n-1}\right)
   \leftrightarrow\left(\alpha_{2n^{\prime}},Z_{2n^{\prime}}\right)$.
Therefore the combination (\ref{eq:combination}) transforms in the
same way as the correlation function (\ref{eq:SFTcorr}) under the
permutation $r\leftrightarrow s$.
Namely, the combination~(\ref{eq:combination}) has the right properties 
to be used to express 
the amplitude obtained from
the correlation function~(\ref{eq:SFTcorr}).

We would like to arrange the three-string vertex so that the worldsheet
correlation functions always appear in the form~(\ref{eq:combination}).
This can be achieved by defining the three-string 
vertex~(\ref{eq:3vertex2})
which involves fermions as 
\begin{eqnarray}
\lefteqn{
   \int dt\int d1d2d3
      \left\langle V_{3}\left(1,2,3\right)|\Phi(t)\right\rangle _{1}
      \left|\Phi(t)\right\rangle _{2}\left|\Phi(t)\right\rangle _{3}}
\nonumber \\
 &  & =\int\prod_{r=1}^{3}
    \left(\frac{d^{d}p_{r}}{\left(2\pi\right)^{d}}\right)
    \left(2\pi\right)^{d}\delta^{d}
    \Biggl(\sum_{r=1}^{3}p_{r}\Biggr)
    e^{-\Gamma^{\left[3\right]}\left(1,2,3\right)}
    \alpha_{3}
    f\left(\alpha_{1},\alpha_{2},\alpha_{3};Z_{1},Z_{2},Z_{3}\right)
\nonumber \\
 &  & \hphantom{=\prod\int}
     \qquad\qquad\,\times
     \left\langle V_{3}^{\mathrm{LPP}}\left(1,2,3\right)\right|
         P_{123}
      \Bigl|\tilde{\Phi}\left(p_{1}^{-}\right)\Bigr\rangle_{1}
      \Bigl|\tilde{\Phi}\left(p_{2}^{-}\right)\Bigr\rangle_{2}
      \Bigl|\tilde{\Phi}\left(p_{3}^{-}\right)\Bigr\rangle_{3}~,
\end{eqnarray}
 where $\left\langle V_{3}^{\mathrm{LPP}}\left(1,2,3\right)\right|P_{123}$
is defined using the correlation functions of the vertex operators
at $z=Z_{r}\,\left(r=1,2,3\right)$ and with cuts in Fig.~\ref{fig:cuts}.
Then it is straightforward to show that amplitudes can be given by
the integral of eq.(\ref{eq:combination}) by checking the factorization 
properties.
Since the combination~(\ref{eq:combination}) is specified by 
$Z_{r},\bar{Z}_{r}$
and information on the external lines, we can see that contributions
from various channels to the amplitudes are smoothly connected. Thus
a tree-level amplitude can be given as an integral over the moduli
space.

\subsubsection*{LPP vertex $V_{3}^{\mathrm{LPP}}$}

\begin{figure}[h]
\begin{centering}
\includegraphics[scale=0.5]{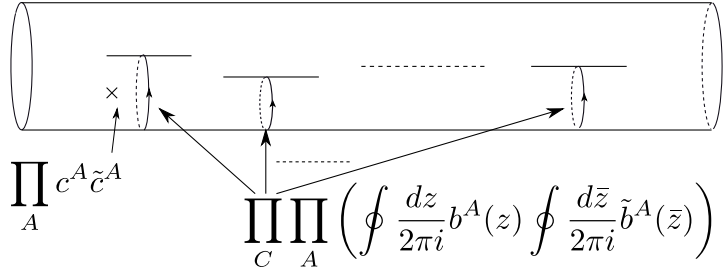} 
\par\end{centering}

\caption{Insertions of $b^{A},c^{A},\tilde{b}^{A},\tilde{c}^{A}$
          \label{fig:insertions-of}}

\end{figure}

In order to define $V_{3}^{\mathrm{LPP}}$ for the worldsheet theory
in section~\ref{sec:Dimensional-regularization}, we need to specify
the way to treat the zero modes of the ghost-like variables. We define
the LPP vertex so that 
\begin{equation}
\left\langle V_{3}^{\mathrm{LPP}}\left(1,2,3\right)\right|
   \mathcal{O}
  \left|0\right\rangle _{1}\left|0\right\rangle _{2}
  \left|0\right\rangle _{3}
=\left\langle \prod_{A}
         \left(c^{A}\left(z_{0}\right)
                \tilde{c}^{A}\left(\bar{z}_{0}\right)\right)
         \mathcal{O}
  \right\rangle _{\mathrm{LC}}\,,
\label{eq:V3LPP}
\end{equation}
 for $\mathcal{O}$ which depends on $c^{A},\tilde{c}^{A}$ only through
their derivatives. Since the string field $\left|\Phi\right\rangle $
satisfies the condition~(\ref{eq:b0A}), 
eq.(\ref{eq:V3LPP}) for such $\mathcal{O}$
is enough to define the string field action. 
%It is straightforward
%to construct the correlation functions using the definition of the
%propagator and the vertex (\ref{eq:V3LPP}). 
With the condition~(\ref{eq:b0A})
and the vertex (\ref{eq:V3LPP}), the correlation functions on the
worldsheet are with insertions of 
$\prod_{A}\left(c^{A}\left(z_{0}\right)
                \tilde{c}^{A}\left(\bar{z}_{0}\right)\right)$
and 
$\prod_{A,C}\left(\oint_{C}\frac{dz}{2\pi i}b^{A}\left(z\right)
                  \oint_{C}\frac{d\bar{z}}{2\pi i}\tilde{b}^{A}
                        \left(\bar{z}\right)\right)$,
which soak up the zero modes. The contours $C$ are taken to be the
ones depicted in Fig.~\ref{fig:insertions-of} on light-cone diagrams
for multi-loop amplitudes. 
The insertion of 
$\prod_{A} \left(c^{A}\left(z_{0}\right)
                 \tilde{c}^{A}\left(\bar{z}_{0}\right)\right)$ 
soaks up the zero modes of $c^A,\tilde{c}^A$ with weight $(0,0)$ and 
it works as the insertion of $\xi$ in the bosonized superghost system. 
The condition that the operator $\mathcal{O}$ depends on 
$c^{A},\tilde{c}^{A}$ 
only through their derivatives means 
that it is in the ``small Hilbert space", 
in which the modes $c_0^A,~\tilde{c}_0^A$ are absent. 
Therefore the right hand side of eq.(\ref{eq:V3LPP}) 
can be expressed as 
\begin{equation}
\left\langle 
         \mathcal{O}
  \right\rangle _{\mathrm{small~Hilbert~space}}~.
\label{eq:small}
\end{equation}

The Fock vacua of 
$\beta^{A},\gamma^{A},\tilde{\beta}^{A},\tilde{\gamma}^{A}$
are defined so that the positive frequency modes annihilate them. 
There exist zero modes on the worldsheet 
when the spin structure is odd and the correlation
functions become infinite. This will be a problem in calculating multi-loop
amplitudes as is mentioned in section~\ref{sec:Conclusions-and-discussions}.

Putting all the above results together, we obtain the tree-level
$N$-string amplitudes $\mathcal{A}_{N}$ in eq.(\ref{eq:AN-1}).

\bibliographystyle{utphys} 
\bibliography{DR,SFTMar8_11}

\providecommand{\href}[2]{#2}\begingroup\raggedright\begin{thebibliography}{10}

\bibitem{Greensite:1986gv}
J.~Greensite and F.~R. Klinkhamer, ``{NEW INTERACTIONS FOR SUPERSTRINGS},''
\href{http://dx.doi.org/10.1016/0550-3213(87)90256-2}{{\em Nucl. Phys.} {\bf
  B281} (1987)  269}.
%%CITATION = NUPHA,B281,269;%%.

\bibitem{Greensite:1987hm}
J.~Greensite and F.~R. Klinkhamer, ``{SUPERSTRING AMPLITUDES AND CONTACT
  INTERACTIONS},''
\href{http://dx.doi.org/10.1016/0550-3213(88)90622-0}{{\em Nucl. Phys.} {\bf
  B304} (1988)  108}.
%%CITATION = NUPHA,B304,108;%%.

\bibitem{Greensite:1987sm}
J.~Greensite and F.~R. Klinkhamer, ``{CONTACT INTERACTIONS IN CLOSED
  SUPERSTRING FIELD THEORY},''
\href{http://dx.doi.org/10.1016/0550-3213(87)90485-8}{{\em Nucl. Phys.} {\bf
  B291} (1987)  557}.
%%CITATION = NUPHA,B291,557;%%.

\bibitem{Green:1987qu}
M.~B. Green and N.~Seiberg, ``{CONTACT INTERACTIONS IN SUPERSTRING THEORY},''
\href{http://dx.doi.org/10.1016/0550-3213(88)90549-4}{{\em Nucl. Phys.} {\bf
  B299} (1988)  559}.
%%CITATION = NUPHA,B299,559;%%.

\bibitem{Wendt:1987zh}
C.~Wendt, ``{SCATTERING AMPLITUDES AND CONTACT INTERACTIONS IN WITTEN'S
  SUPERSTRING FIELD THEORY},''
\href{http://dx.doi.org/10.1016/0550-3213(89)90118-1}{{\em Nucl. Phys.} {\bf
  B314} (1989)  209}.
%%CITATION = NUPHA,B314,209;%%.

\bibitem{Baba:2009kr}
Y.~Baba, N.~Ishibashi, and K.~Murakami, ``{Light-Cone Gauge Superstring Field
  Theory and Dimensional Regularization},''
  \href{http://dx.doi.org/10.1088/1126-6708/2009/10/035}{{\em JHEP} {\bf 10}
  (2009)  035},
\href{http://arxiv.org/abs/0906.3577}{{\tt arXiv:0906.3577 [hep-th]}}.
%%CITATION = 0906.3577;%%.

\bibitem{Baba:2009ns}
Y.~Baba, N.~Ishibashi, and K.~Murakami, ``{Light-Cone Gauge String Field Theory
  in Noncritical Dimensions},''
  \href{http://dx.doi.org/10.1088/1126-6708/2009/12/010}{{\em JHEP} {\bf 12}
  (2009)  010},
\href{http://arxiv.org/abs/0909.4675}{{\tt arXiv:0909.4675 [hep-th]}}.
%%CITATION = 0909.4675;%%.

\bibitem{Baba:2009fi}
Y.~Baba, N.~Ishibashi, and K.~Murakami, ``{Light-cone Gauge NSR Strings in
  Noncritical Dimensions},''
  \href{http://dx.doi.org/10.1007/JHEP01(2010)119}{{\em JHEP} {\bf 01} (2010)
  119},
\href{http://arxiv.org/abs/0911.3704}{{\tt arXiv:0911.3704 [hep-th]}}.
%%CITATION = 0911.3704;%%.

\bibitem{Baba:2009zm}
Y.~Baba, N.~Ishibashi, and K.~Murakami, ``{Light-cone Gauge Superstring Field
  Theory and Dimensional Regularization II},''
  \href{http://dx.doi.org/10.1007/JHEP08(2010)102}{{\em JHEP} {\bf 08} (2010)
  102},
\href{http://arxiv.org/abs/0912.4811}{{\tt arXiv:0912.4811 [hep-th]}}.
%%CITATION = 0912.4811;%%.

\bibitem{Mandelstam:1973jk}
S.~Mandelstam, ``{Interacting String Picture of Dual Resonance Models},''
\href{http://dx.doi.org/10.1016/0550-3213(73)90622-6}{{\em Nucl. Phys.} {\bf
  B64} (1973)  205--235}.
%%CITATION = NUPHA,B64,205;%%.

\bibitem{Kaku:1974zz}
M.~Kaku and K.~Kikkawa, ``{The Field Theory of Relativistic Strings, Pt. 1.
  Trees},''
\href{http://dx.doi.org/10.1103/PhysRevD.10.1110}{{\em Phys. Rev.} {\bf D10}
  (1974)  1110}.
%%CITATION = PHRVA,D10,1110;%%.

\bibitem{Kaku:1974xu}
M.~Kaku and K.~Kikkawa, ``{The Field Theory of Relativistic Strings. 2. Loops
  and Pomerons},''
\href{http://dx.doi.org/10.1103/PhysRevD.10.1823}{{\em Phys. Rev.} {\bf D10}
  (1974)  1823--1843}.
%%CITATION = PHRVA,D10,1823;%%.

\bibitem{Cremmer:1974ej}
E.~Cremmer and J.-L. Gervais, ``{Infinite Component Field Theory of Interacting
  Relativistic Strings and Dual Theory},''
\href{http://dx.doi.org/10.1016/0550-3213(75)90655-0}{{\em Nucl. Phys.} {\bf
  B90} (1975)  410--460}.
%%CITATION = NUPHA,B90,410;%%.

\bibitem{Mandelstam:1974hk}
S.~Mandelstam, ``{Interacting String Picture of the Neveu-Schwarz-Ramond
  Model},''
\href{http://dx.doi.org/10.1016/0550-3213(74)90127-8}{{\em Nucl. Phys.} {\bf
  B69} (1974)  77--106}.
%%CITATION = NUPHA,B69,77;%%.

\bibitem{Sin:1988yf}
S.-J. Sin, ``{GEOMETRY OF SUPER LIGHT CONE DIAGRAMS AND LORENTZ INVARIANCE OF
  LIGHT CONE STRING FIELD THEORY. 2. CLOSED NEVEU-SCHWARZ STRING},''
\href{http://dx.doi.org/10.1016/0550-3213(89)90517-8}{{\em Nucl. Phys.} {\bf
  B313} (1989)  165}.
%%CITATION = NUPHA,B313,165;%%.

\bibitem{Ishibashi:2010nq}
N.~Ishibashi and K.~Murakami, ``{Light-cone Gauge NSR Strings in Noncritical
  Dimensions II -- Ramond Sector},''
  \href{http://dx.doi.org/10.1007/JHEP01(2011)008}{{\em JHEP} {\bf 01} (2011)
  008},
\href{http://arxiv.org/abs/1011.0112}{{\tt arXiv:1011.0112 [hep-th]}}.
%%CITATION = 1011.0112;%%.

\bibitem{Friedan:1985ge}
D.~Friedan, E.~J. Martinec, and S.~H. Shenker, ``{Conformal Invariance,
  Supersymmetry and String Theory},''
{\em Nucl. Phys.} {\bf B271} (1986)  93.
%%CITATION = NUPHA,B271,93;%%.

\bibitem{D'Hoker:2002gw}
E.~D'Hoker and D.~H. Phong, ``{Lectures on Two-Loop Superstrings},''
\href{http://arxiv.org/abs/hep-th/0211111}{{\tt arXiv:hep-th/0211111}}.
%%CITATION = HEP-TH/0211111;%%.

\bibitem{'tHooft:1972fi}
G.~'t~Hooft and M.~J.~G. Veltman, ``{Regularization and Renormalization of
  Gauge Fields},''
\href{http://dx.doi.org/10.1016/0550-3213(72)90279-9}{{\em Nucl. Phys.} {\bf
  B44} (1972)  189--213}.
%%CITATION = NUPHA,B44,189;%%.

\bibitem{Akyeampong:1973xi}
D.~A. Akyeampong and R.~Delbourgo, ``{Dimensional regularization, abnormal
  amplitudes and anomalies},''
\href{http://dx.doi.org/10.1007/BF02786835}{{\em Nuovo Cim.} {\bf A17} (1973)
  578--586}.
%%CITATION = NUCIA,A17,578;%%.

\bibitem{Breitenlohner:1977hr}
P.~Breitenlohner and D.~Maison, ``{Dimensional Renormalization and the Action
  Principle},''
\href{http://dx.doi.org/10.1007/BF01609069}{{\em Commun. Math. Phys.} {\bf 52}
  (1977)  11--38}.
%%CITATION = CMPHA,52,11;%%.

\bibitem{Kugo:1992md}
T.~Kugo and B.~Zwiebach, ``{Target space duality as a symmetry of string field
  theory},'' \href{http://dx.doi.org/10.1143/PTP.87.801}{{\em Prog. Theor.
  Phys.} {\bf 87} (1992)  801--860},
\href{http://arxiv.org/abs/hep-th/9201040}{{\tt arXiv:hep-th/9201040}}.
%%CITATION = HEP-TH/9201040;%%.

\bibitem{LeClair:1988sp}
A.~LeClair, M.~E. Peskin, and C.~R. Preitschopf, ``{String Field Theory on the
  Conformal Plane. 1. Kinematical Principles},''
\href{http://dx.doi.org/10.1016/0550-3213(89)90075-8}{{\em Nucl. Phys.} {\bf
  B317} (1989)  411}.
%%CITATION = NUPHA,B317,411;%%.

\bibitem{LeClair:1988sj}
A.~LeClair, M.~E. Peskin, and C.~R. Preitschopf, ``{String Field Theory on the
  Conformal Plane. 2. Generalized Gluing},''
\href{http://dx.doi.org/10.1016/0550-3213(89)90076-X}{{\em Nucl. Phys.} {\bf
  B317} (1989)  464}.
%%CITATION = NUPHA,B317,464;%%.

\end{thebibliography}\endgroup

\end{document}